\documentclass[preprint,showpacs,preprintnumbers,amsmath,amssymb,floats,amsfonts,superscriptaddress,nofootinbib,11pt]{revtex4-2}
\usepackage[hidelinks]{hyperref}
\usepackage{graphicx}
\usepackage{tabularx}
\usepackage{siunitx}
\usepackage{epstopdf} 

\begin{document}

\title[Connecting scattering, monodromy, and MST's renormalized angular momentum in Kerr spacetime]{Connecting scattering, monodromy, and MST's renormalized angular momentum for the Teukolsky equation in Kerr spacetime}

\author{Zachary Nasipak}
\affiliation{School of Mathematical Sciences and STAG Research Centre,
University of Southampton, Southampton, SO17 1BJ, United Kingdom}
\email{z.nasipak@soton.ac.uk}
\affiliation{Center for Space Sciences and Technology, University of Maryland Baltimore County, Baltimore, MD, 21250, USA}
\affiliation{Department of Physics and Astronomy, University of North Carolina at Chapel Hill, Chapel Hill, North Carolina 27599}

\begin{abstract}
The Teukolsky equation describes perturbations of Kerr spacetime and is central to the study of rotating black holes and gravitational waves. In the frequency domain, the Teukolsky equation separates into radial and angular ordinary differential equations. Mano, Suzuki, and Takasugi (MST) found semi-analytic solutions to the homogeneous radial Teukolsky equation in terms of series of analytic special functions. The MST expansions hinge on an auxiliary parameter known as the \emph{renormalized angular momentum} $\nu$, which one must calculate to ensure the convergence of these series solutions. In this work, we present a method for calculating $\nu$ via monodromy eigenvalues, which capture the behavior of ordinary differential equations and their solutions in the complex domain near their singular points. We directly relate the monodromy data of the radial Teukolsky equation to the parameter $\nu$ and provide a numerical scheme for calculating $\nu$ based on monodromy. With this method we evaluate $\nu$ in different regions of parameter space and analyze the numerical stability of this approach. We also highlight how, through $\nu$, monodromy data are linked to scattering amplitudes for generic (linear) perturbations of Kerr spacetime.
\end{abstract}

%
%
%
%
%

\maketitle
\newpage 

\section{Introduction}

The Teukolsky equation is a linearized field equation that governs the evolution and propagation of perturbations in a background Kerr spacetime \cite{Teuk73}. From the Teukolsky equation, one can calculate the quasinormal mode ringdowns of merged compact object binaries, the radiative backreaction experienced by small perturbing bodies inspiraling towards rotating black holes, and the gravitational signal radiated by a system and measured by a distant observer. Thus, the Teukolsky equation has been central to the development of gravitational wave science.  

In Boyer-Lindquist coordinates $(t, r, \theta, \phi)$, the Teukolsky equation takes the form,
\begin{multline}
    \label{eqn:teukFull}
    \left[\frac{r^2+a^2}{\Delta} - a^2\sin^2\theta \right] \partial_t^2 \Psi_{s}+\frac{4M a r}{\Delta} \partial_t \partial_\phi \Psi_s + \left[\frac{a^2}{\Delta} - \frac{1}{\sin^2\theta} \right] \partial_\phi^2 \Psi_s 
    \\
    - \Delta^{-s} \partial_r \left(\Delta^{s+1} \partial_r \Psi_s \right)
    - \frac{1}{\sin\theta}\partial_\theta \left(\sin\theta \partial_\theta \Psi_s\right) - 2s\left[\frac{a(r-M)}{\Delta} + i \frac{\cos\theta}{\sin^2\theta} \right] \partial_\phi \Psi_s 
    \\
    - 2s\left[\frac{M(r^2-a^2)}{\Delta} - r - ia\cos\theta \right]\partial_t \Psi_s + (s^2\cot^2\theta - s)\Psi_s = 4\pi \Sigma {T}_s, 
\end{multline}
where $M$ and $a$ are the Kerr mass and spin parameters, $\Delta = r^2-2Mr +a^2$, $\Sigma = r^2 + a^2\cos^2\theta$, $s$ is the spin-weight of the perturbing field $\Psi_s$, and ${T}_s$ is the source of the perturbation. (See Table I in \cite{Teuk73} for exact definitions of $\Psi_s$ and ${T}_s$.) By altering the spin-weight parameter, the Teukolsky equation can describe scalar ($s=0$), neutrino ($s=\pm \frac{1}{2}$), electromagnetic ($s=\pm 1$), and gravitational ($s=\pm2$) perturbations of rotating Kerr black holes. In this work, we focus our attention on the vacuum case of $T_s = 0$.

The Teukolsky equation is amenable to separation of variables in the frequency-domain via the mode decomposition $\Psi_s = \psi_{slm\omega}(t,r,\theta,\phi) = R_{slm\omega}(r) S_{slm\omega}(\theta)e^{im\phi} e^{-i\omega t}$ \cite{BrilETC72, Teuk72}. With this ansatz, Eq.~\eqref{eqn:teukFull} decouples into two ordinary differential equations (ODEs),
\begin{align} \label{eqn:teuk}
	\Delta^{-s}\frac{d}{dr}&\left( \Delta^{s+1}\frac{dR_{slm\omega}}{dr} \right)
	+ \left(\frac{K^2-2is(r-M)K}{\Delta}+{4is\omega r+\lambda^T_{slm\omega}} \right)R_{slm\omega}=0,
 \\ \label{eqn:swsh}
    \frac{d^2S_{slm\omega}}{d\theta^2}\; + \; &
	\frac{\cos\theta}{\sin\theta}\frac{dS_{slm\omega}}{d\theta}
 \\ \notag
 & \qquad \qquad
	- \left(a^2\omega^2\sin^2\theta+\frac{(m+s\cos\theta)^2}{\sin^2\theta}
	+2a\omega s\cos\theta-s-2ma\omega-\lambda^T_{slm\omega} \right)S_{slm\omega} = 0,
\end{align}
where $K=(r^2+a^2)\omega-ma$, and $\lambda^T_{slm\omega}$ is the spheroidal eigenvalue (or separation constant). Solutions to Eq.~\eqref{eqn:swsh} are known as spin-weighted spheroidal harmonics, which are generalizations of the spin-weighted \emph{spherical} harmonics ${}_s Y_{lm}(\theta, \phi)$. For $a\omega \rightarrow 0$, the two sets of harmonics are equivalent, with $S_{slm\omega}(\theta)e^{im\phi} \rightarrow {}_s Y_{lm}(\theta, \phi)$ and $\lambda^T_{slm\omega} \rightarrow l(l+1) - s(s+1)$. The numerical calculation of $S_{slm\omega}(\theta)$ is well understood \cite{Leav86, Hugh00b}, and several open-source tools are available for producing these harmonics (e.g., \cite{BHPT_SWSH, BHPT_KGW, Park_SPH, Lo_SWSH}).

The construction of the homogeneous radial solutions $R_{slm\omega}$ has also been extensively studied in the literature \cite{SasaNaka82, Leav86, ManoSuzuTaka96b, Hugh00, FujiTago04, FujiTago05}. Mano, Sukuzi, and Takasugi (MST) devised one commonly used method \cite{ManoSuzuTaka96b}, in which the homogeneous radial solutions are represented as semi-analytic series of hypergeometric functions. To evaluate these series, one must first solve for an auxiliary parameter known as the \emph{renormalized angular parameter} $\nu$, which controls whether or not each series solution converges. Consequently, finding a ``convergent'' value of $\nu$ is critical to solving the Teukolsky equation with series of analytic functions.

MST found that their series converge when $\nu$ is a root of a particular three-term continued fraction equation (see Eq.~(123) in Ref.~\cite{SasaTago03}). Therefore, many researchers have relied on sophisticated root-finding methods to numerically determine $\nu$ \cite{FujiTago05, Thro10}, but these procedures can struggle at high frequencies ($\omega > 1$) and for large values of the spheroidal mode number $l$, particularly when $\nu$ becomes complex. Alternatively, recent codes (e.g., \cite{BHPT_TEUK}) have employed a new algorithm inspired by the work of Castro et al.~\cite{CastETC13a, CastETC13b}, which determines $\nu$ based on the \emph{monodromy} data that capture the behavior of the radial Teukolsky solutions as they circle the irregular singular point at $r=\infty$. 

While this {monodromy} approach has proven to be highly successful, there is little written about its application to the MST solutions. Ref.~\cite{KavaOtteWard16} hints at the connection between $\nu$ and monodromy theory in their post-Newtonian expansions of the MST solutions, while Refs.~\cite{CasaZimm19,CasaMicc19} identify the connection between $\nu$ and the monodromy eigenvalues of the Teukolsky equation, but they do not provide an exact relationship. Ref.~\cite{BautETC24} found that their gauge modulus parameter $a$ (which is directly related to monodromy eigenvalues) satisfies the relation $a = -\nu- 1/2$, but this result was only verified up to 9th post-Minkowskian order. Refs.~\cite{CarnNova16,NovaETC19} also found a relationship between $\nu$ and monodromy data when extending the MST solutions to the Teukolsky equation in Kerr-de Sitter spacetime; however, these results were not generalized to Kerr. In this work, we derive an exact relationship between the monodromy eigenvalues of the Teukolsky equation in Kerr spacetime (for arbitrary values of the spin-weight $s$) and the renormalized angular momentum $\nu$ used in the MST solutions. We also provide numerical methods for calculating $\nu$ based on monodromy theory.

\subsection{Paper outline}

In Section \ref{sec:series} we review both asymptotic and MST series solutions to the homogeneous radial Teukolsky equation in Kerr spacetime. In Section \ref{sec:monodromy} we summarize the works of \cite{CastETC13a, CastETC13b, DaalOlve95}, which outline how monodromy methods are used to describe the solutions of ODEs as they ``run around'' singular points in the complex domain. As an example, we apply these methods to determine the monodromy eigenvalues of confluent hypergeometric functions. In Section \ref{sec:nuFromMono} we use these results to connect the monodromy eigenvalues of the Teukolsky equation to the renormalized angular momentum $\nu$. We then present new numerical methods for calculating the monodromy eigenvalues of the Teukolsky equation and $\nu$ in Section \ref{sec:numerical}. We also highlight the numerical advantages and limitations of solving for $\nu$ via monodromy methods. We end with a discussion of our results in Section \ref{sec:conclusion}. For this paper we use the metric signature $(-+++)$, the sign conventions, where applicable, of \cite{MisnThorWhee73}, and units such that $c=G=1$.

\section{Series solutions of the radial Teukolsky equation}
\label{sec:series}

It is often advantageous to characterize solutions of an ODE in terms of the equation's singular points. The radial Teukolsky equation possesses three: two regular singular points at the inner and outer horizons $r_\pm = M \pm \sqrt{M^2-a^2}$ and one irregular singular point (of Poincar\'e rank one) at infinity. In this work we primarily focus on homogeneous radial solutions that are defined on the physical domain $r \in [r_+, \infty]$ and, consequently, by their behavior at the points $r=r_+$ and $r=\infty$. For radiative modes ($\omega \neq 0$), four common solutions are,
\begin{subequations} \label{eqn:teukAsymp}
    \begin{align}
    R^\mathrm{in}_{slm\omega}(r\rightarrow r_+) &\sim \mathcal{R}^\mathrm{in,trans}_{slm\omega} \Delta^{-s} e^{-ikr_*},
    &
    R^\mathrm{up}_{slm\omega}(r\rightarrow \infty) &\sim \mathcal{R}^\mathrm{up,trans}_{slm\omega} r^{-(2s+1)}e^{i\omega r_*},
    \\
    R^\mathrm{out}_{slm\omega}(r\rightarrow r_+) &\sim \mathcal{R}^\mathrm{out,trans}_{slm\omega} e^{ikr_*},
    &
    R^\mathrm{down}_{slm\omega}(r\rightarrow \infty) &\sim \mathcal{R}^\mathrm{down,trans}_{slm\omega} r^{-1}e^{-i\omega r_*},
\end{align}
\end{subequations}
where $k = \omega - m \Omega_+$, $\Omega_+ = a/(2Mr_+)$, and $r_*$ is the tortoise coordinate defined by the differential relation $dr_*/dr = (r^2+a^2)/\Delta$. For scattering problems, it is also useful to consider the additional asymptotic behaviors,
\begin{subequations} \label{eqn:teukAsymp2}
   \begin{align}
    R^\mathrm{in}_{slm\omega}(r\rightarrow \infty) &\sim \mathcal{R}^\mathrm{in,ref}_{slm\omega} r^{-(2s+1)}{e^{i\omega r_*}} + \mathcal{R}^\mathrm{in,inc}_{slm\omega} r^{-1} {e^{-i\omega r_*}},
    \\
    R^\mathrm{out}_{slm\omega}(r\rightarrow \infty) &\sim \mathcal{R}^\mathrm{out,inc}_{slm\omega} r^{-(2s+1)}{e^{i\omega r_*}} + \mathcal{R}^\mathrm{out,ref}_{slm\omega} r^{-1} {e^{-i\omega r_*}},
    \\
    R^\mathrm{up}_{slm\omega}(r\rightarrow r_+) &\sim \mathcal{R}^\mathrm{up,inc}_{slm\omega} e^{ik r_*} + \mathcal{R}^\mathrm{up,ref}_{slm\omega} \Delta^{-s} e^{-ik r_*},
    \\
    R^\mathrm{down}_{slm\omega}(r\rightarrow r_+) &\sim \mathcal{R}^\mathrm{down,ref}_{slm\omega} e^{ik r_*} + \mathcal{R}^\mathrm{down,inc}_{slm\omega} \Delta^{-s} e^{-ik r_*},
\end{align} 
\end{subequations}
where $\mathcal{R}^\mathrm{A,trans}_{slm\omega}$, $\mathcal{R}^\mathrm{A,inc}_{slm\omega}$, and $\mathcal{R}^\mathrm{A,ref}_{slm\omega}$ are transmission, incidence, and reflection scattering amplitudes for A = \{in, up, out, down\}.

In the remainder of this section, we review different series solutions of the radial Teukolsky equation for $\omega \neq 0$ modes. First we outline series expansions around the singular points of the Teukolsky equation, $r=r_+$ and $r=\infty$, and the simplification of these series in confluent Heun form. We then summarize the semi-analytic series solutions provided by MST. To condense notation, we define the dimensionless parameters,
\begin{align*}
    \chi &= a / M,
    &
    \kappa &= \sqrt{1 - \chi^2},
    &
    \epsilon &= 2M \omega,
    \\
    \tau &= (\epsilon - m\chi)/\kappa,
    & 
    \xi &= s - i\epsilon,
    &
    \epsilon_\pm &= (\epsilon \pm \tau)/2,
\end{align*}
which will be used throughout the rest of this paper.

\subsection{Frobenius-Fuch and asymptotic series}

Because $r=r_+$ is a regular singular point, $R^\mathrm{in}$ and $R^\mathrm{out}$ can be approximated by the Frobenius-Fuch series,
\begin{subequations}\label{eqn:horizonSeries}
    \begin{align}
        R^\mathrm{in}(r\rightarrow r_+) \simeq R^\mathcal{H}_2(r) &= (r-r_+)^{-s-i\epsilon_+} \sum_{n=0}^\infty a_{2,n} {(r-r_+)^n},
        \\
        R^\mathrm{out}(r\rightarrow r_+) \simeq R^\mathcal{H}_1(r) &= (r-r_+)^{i\epsilon_+} \sum_{n=0}^\infty a_{1,n} {(r-r_+)^n},
    \end{align}
\end{subequations}
where we have suppressed the $(slm\omega)$ subscripts for brevity. On the other hand, $R^\mathrm{up}$ and $R^\mathrm{down}$ are typically approximated in terms of the asymptotic series,
\begin{subequations} \label{eqn:infinitySeries}
    \begin{align}
        R^\mathrm{up}(r \rightarrow \infty) &\sim R^\mathcal{I}_1(r) = e^{i\omega r} r^{-1-2s + i\epsilon} \sum_{n=0}^\infty b_{1,n} r^{-n},
        \\
        R^\mathrm{down}(r \rightarrow \infty) &\sim R^\mathcal{I}_2(r) = e^{-i\omega r} r^{-1-i\epsilon} \sum_{n=0}^\infty b_{2,n} r^{-n}.
    \end{align}  
\end{subequations}
The horizon series \eqref{eqn:horizonSeries} have radii of convergence $|r-r_+| < 2M\sqrt{1-\chi^2}$, while the infinity series \eqref{eqn:infinitySeries} are formally non-convergent, making both expansions poor representations for much of the radial domain. Nonetheless these series are particularly useful for numerically approximating solutions near the boundaries, providing important initial data for numerical ODE solvers. 

The calculation of these series expansions is further simplified by putting the Teukolsky equation into confluent Heun form,
\begin{equation} \label{eqn:che}
    \frac{d^2w}{d\hat{z}^2} + \left(\frac{\gamma_\mathrm{CH}}{\hat{z}} + \frac{\delta_\mathrm{CH}}{\hat{z}-1} + \varepsilon_\mathrm{CH} \right)
    \frac{dw}{d\hat{z}} + \frac{\alpha_\mathrm{CH} \hat{z} - q_\mathrm{CH}}{\hat{z}(\hat{z}-1)} w = 0,
\end{equation}
via the transformations,
\begin{align} \label{eqn:cheTransform}
    \epsilon \kappa \hat{z} = \omega( r- r_-),
    \qquad \qquad
    \epsilon \kappa (\hat{z} - 1) = \omega( r- r_+),
    \qquad \qquad
    R(\hat{z}) = \hat{z}^{a} (\hat{z}-1)^{b} e^{c \hat{z}} w(\hat{z}),
\end{align}
where
\begin{align} \label{eqn:abTransform}
    2a &= -s + n_a(s + 2i\epsilon_-),
    &
    2b &= -s + n_b(s + 2i\epsilon_+),
    &
    c &= i n_c\epsilon \kappa,
\end{align}
$n_a = \pm 1$, $n_b = \pm 1$, $n_c = \pm 1$, and the confluent Heun parameters are given by
\begin{subequations}
    \begin{gather}
    	\gamma_\mathrm{CH} = 1+s+2a, \qquad
    	\delta_\mathrm{CH} = 1+s+2b, \qquad
    	\varepsilon_\mathrm{CH} = 2c,
        \\
    	\alpha_\mathrm{CH} = 2c\left(1 +s+a+b\right) + 2i\epsilon\kappa \xi,
    	\\
    	q_\mathrm{CH} = -(a+b+c)(s+1) - 2a b + \lambda
    	- 2\epsilon_+ \epsilon_- + \epsilon[ m\chi
    	-i \xi (1-\kappa)] + 2ac.
    \end{gather} 
\end{subequations}
Note that this transformation is not unique. Due to our freedom in choosing $n_a$, $n_b$, and $n_c$, Eq.~\eqref{eqn:abTransform} provides eight different combinations of $a$, $b$, and $c$ that will transform the Teukolsky solutions into solutions of Eq.~\eqref{eqn:che}. In the remainder of this section, we will make use of different transformation choices when examining the asymptotic behavior of $w(z)$.

Eq.~\eqref{eqn:che} has singular points at $\hat{z}=\{0, 1, \infty\}$, with the latter two corresponding to the relevant physical boundaries at the horizon and infinity. Near the horizon, solutions take the asymptotic forms,
\begin{subequations} \label{eqn:cheRegular}
    \begin{align}
        w^\mathrm{in}(\hat{z}\rightarrow 1) &\simeq w^\mathcal{H}_2(\hat{z}) = (\hat{z} - 1)^{\lambda^\mathcal{H}_2} \sum_{k=0}^\infty \hat{a}_{2,k} (\hat{z} - 1)^k,
        \\
        w^\mathrm{out}(\hat{z}\rightarrow 1) &\simeq w^\mathcal{H}_1(\hat{z}) = (\hat{z} - 1)^{\lambda^\mathcal{H}_1} \sum_{k=0}^\infty \hat{a}_{1,k} (\hat{z} - 1)^k,
    \end{align}
\end{subequations}
with indices $\lambda^\mathcal{H}_1 = 0$ and $\lambda^\mathcal{H}_2 = 1 - \delta_\mathrm{CH}$, and the coefficients satisfy the three-term recurrence relation,
\begin{equation} \label{eqn:cheRegRecurrence}
    {A}^\mathcal{H}_{j,k} \hat{a}_{j,k - 1} + {B}^\mathcal{H}_{j,k} \hat{a}_{j,k} + {C}^\mathcal{H}_{j,k} \hat{a}_{j,k + 1} = 0.
\end{equation}
A common choice of initial conditions is $\hat{a}_{j,0} = 1$, and $\hat{a}_{j,-1} = 0$. See Appendix \ref{app:CHEdefinitions} for exact definitions of ${A}^\mathcal{H}_{j,k}$, ${B}^\mathcal{H}_{j,k}$, and ${C}^\mathcal{H}_{j,k}$. Choosing $(n_a, n_b, n_c) = (-1, +1, -1)$, $w^\mathrm{in}$ and $w^\mathrm{out}$ transform to $R^\mathrm{in}$ and $R^\mathrm{out}$, respectively, via Eq.~\eqref{eqn:cheTransform}.\footnote{Meanwhile, switching the sign of $n_b$ swaps this relationship, so that $w^\mathrm{in}$ and $w^\mathrm{out}$ transform to $R^\mathrm{out}$ and $R^\mathrm{in}$, respectively. Changing the signs of $n_c$ and $n_a$ simply affect the overall normalization of the solutions.}

Near infinity we have,
\begin{subequations} \label{eqn:cheIrregular}
    \begin{align}
        w^\mathrm{up}(\hat{z}\rightarrow \infty) &\sim w^\mathcal{I}_1(\hat{z}) = e^{\mu^\mathcal{I}_1 \hat{z}} \hat{z}^{\lambda^\mathcal{I}_1} \sum_{k=0}^\infty \hat{b}_{1,k} \hat{z}^{-k},
        \\
        w^\mathrm{down}(\hat{z}\rightarrow \infty) &\sim w^\mathcal{I}_2(\hat{z}) = e^{\mu^\mathcal{I}_2 \hat{z}} \hat{z}^{\lambda^\mathcal{I}_2} \sum_{k=0}^\infty \hat{b}_{2,k} \hat{z}^{-k},
    \end{align}
\end{subequations}
where $\mu^\mathcal{I}_1 = 0$, $\mu^\mathcal{I}_2 = -\varepsilon_\mathrm{CH}$, $\lambda^\mathcal{I}_1 = -\alpha_\mathrm{CH}/\varepsilon_\mathrm{CH}$, and $\lambda^\mathcal{I}_2 = \alpha_\mathrm{CH}/\varepsilon_\mathrm{CH} - \gamma_\mathrm{CH} - \delta_\mathrm{CH}$; and the coefficients satisfy the three-term recurrence relation,
\begin{equation} \label{eqn:cheRecurrence}
    {A}^\mathcal{I}_{j,k} \hat{b}_{j,k - 1} + {B}^\mathcal{I}_{j,k} \hat{b}_{j,k} + {C}^\mathcal{I}_{j,k} \hat{b}_{j,k + 1} = 0,
\end{equation}
with $\hat{b}_{j,0} = 1$, and $\hat{b}_{j,-1} = 0$. Again, see Appendix \ref{app:CHEdefinitions} for the forms of ${A}^\mathcal{I}_{j,k}$, ${B}^\mathcal{I}_{j,k}$, and ${C}^\mathcal{I}_{j,k}$. Choosing $(n_a, n_b, n_c) = (+1, +1, +1)$, $w^\mathrm{up}$ and $w^\mathrm{down}$ transform to $R^\mathrm{up}$ and $R^\mathrm{down}$, respectively, via Eq.~\eqref{eqn:cheTransform}.\footnote{Changing the sign of $n_c$ leads to $w^\mathrm{up}$ and $w^\mathrm{down}$ transforming to $R^\mathrm{down}$ and $R^\mathrm{up}$, respectively. Altering the signs of $n_a$ and $n_b$ does not affect the asymptotic relationship at infinity.} 

\subsection{MST series}
\label{sec:MST}

The MST Teukolsky solutions are defined in terms of the dimensionless variables,
\begin{align} \label{eqn:MSTcoord}
    x &= \frac{r_+-r}{2M\kappa},
    &
    z &=  \omega (r - r_-),
\end{align}
leading to the series expansions,
\begingroup
\allowdisplaybreaks
\begin{subequations}
    \begin{align} \label{eqn:Rin}
        R^{\mathrm{in}}(x) & =e^{i \epsilon \kappa x}(-x)^{-s-i\epsilon_+}(1-x)^{i\epsilon_-}
        \\
        & \qquad  \times \sum_{n=-\infty}^{\infty} f_{n}^{\nu} \, {}_{2}F_{1}(n+\nu+1-i\tau,\, -n-\nu-i\tau;\, 1-\bar{\xi}-i\tau;\, x), \notag
        \\
        \label{eqn:Rout}
        R^{\mathrm{out}}(x) & =e^{i \epsilon \kappa x}(-x)^{i\epsilon_+}(1-x)^{-s-i\epsilon_-}
        \\
        & \qquad  \times \sum_{n=-\infty}^{\infty} \frac{(\nu+1+i\tau)_n (\nu+1+\bar{\xi})_n}{(\nu+1-i\tau)_n (\nu+1-\bar{\xi})_n} f_{n}^{\nu} \, {}_{2}F_{1}(n+\nu+1+i\tau,\, -n-\nu+i\tau;\, 1+\bar{\xi}+i\tau;\, x), \notag
        \\ \label{eqn:Rup}
        R^{\mathrm{up}}( z) & =  2^{\nu} e^{-i \pi(\nu+1+\xi)} e^{i z} z^{\nu+i\epsilon_+}(z-\epsilon \kappa)^{-s-i\epsilon_+}
        \\
        & \qquad \times \sum_{n=-\infty}^{\infty} \frac{(\nu + 1 + \xi)_{n}}{(\nu + 1 - \xi)_{n}} f_{n}^{\nu}(2iz)^{n} U(n + \nu + 1 + \xi,\, 2n + 2\nu + 2 ;\, -2i{z}), \notag
        \\ \label{eqn:Rdown}
        R^{\mathrm{down}}( z) & =  2^{\nu} e^{i \pi(\nu+1-\xi)} e^{-i z} z^{\nu+i\epsilon_+}(z-\epsilon \kappa)^{-s-i\epsilon_+} \frac{\Gamma(\nu+1-\xi)}{\Gamma(\nu+1+\xi)}
        \\
        & \qquad \times \sum_{n=-\infty}^{\infty} f_{n}^{\nu}(2iz)^{n} U(n + \nu + 1 - \xi,\, 2n + 2\nu + 2 ;\, 2i{z}), \notag
    \end{align}
\end{subequations}
\endgroup
where $_2F_1(a,b,c;x)$ is the Gauss hypergeometric function, $U(a,b;z)$ is the irregular confluent hypergeometric function, $f^\nu_n$ are series coefficients (to be further defined later), and $\nu$ is the aforementioned renormalized angular momentum parameter.
Alternatively, $R^\mathrm{in}$ and $R^\mathrm{out}$ can be expressed by the sums,
\begin{align} \label{eqn:RinToRnu0}
    R^\mathrm{in}(x) &= R^\nu_0(x) + R_0^{-\nu-1}(x),
    &
    R^\mathrm{out}(x) &= B^\nu_0 R^\nu_0(x) + B^{-\nu-1}_0 R_0^{-\nu-1}(x),
\end{align}
where $B^\nu_0$ is a ($\nu$-dependent) constant defined in Appendix \ref{app:MST} and,
\begin{align} \label{eqn:Rnu0}
    R^\nu_0 &= e^{i\epsilon \kappa x} (-x)^{-s-i\epsilon_+}(1-x)^{\nu+i\epsilon_+} \frac{\Gamma(1-\bar{\xi}-i\tau)}{\Gamma(\nu+1-i\tau)\Gamma(\nu+1-\bar{\xi})}
    \\ \notag
    & \;\;\, \times \sum_{n=-\infty}^\infty \frac{\Gamma(2n + 2\nu+1)}{(\nu+1-i\tau)_n(\nu+1-\bar{\xi})_n} f_{n}^{\nu}(1-x)^{n} {}_2F_1\left(-n-\nu-i\tau, -n-\nu-\bar{\xi}; -2n-2\nu; \frac{1}{1-x} \right),
\end{align}
is also a solution to the Teukolsky equation. Likewise, $R^\mathrm{up}$ and $R^\mathrm{down}$ can be expressed as sums of two other independent solutions,
\begin{subequations} \label{eqn:RupToRnu}
    \begin{align} 
        R^\mathrm{up}(z) &= \frac{1}{\sin 2\pi\nu}
        \left[e^{-i\pi(\nu+\xi)}\sin\pi(\nu-\xi)R^\nu_C(z) - i e^{-i\pi\xi} \sin\pi(\nu + \xi) R^{-\nu-1}_C(z) \right],
        \\
        R^\mathrm{down}(z) &= \frac{\sin\pi(\nu+\xi)}{\sin 2\pi\nu}\left[e^{i\pi(\nu-\xi)}R^\nu_C(z) + i e^{-i\pi\xi} R^{-\nu-1}_C(z)\right],
    \end{align} 
\end{subequations}
where,
\begin{align} \label{eqn:RnuC}
    R^\nu_C(z) &= 2^\nu e^{-iz} z^{\nu+i\epsilon_+}\left(z-{\epsilon\kappa}\right)^{-s-i\epsilon_+} \frac{\Gamma(\nu+1-\xi)}{\Gamma(2\nu+2)}
    \\ \notag
    & \qquad \times \sum_{n=-\infty}^\infty \frac{(\nu+1+\xi)_n}{(2\nu +2)_{2n}} f_{n}^{\nu}(-2iz)^{n} M(n+\nu+1-\xi, 2\nu+2n +2; 2iz),
\end{align}
and $M(a,b;z)$ is the confluent hypergeometric function that is regular at $z=0$. [See Sec.~\ref{sec:example} for more details about $M(a, b; z)$ and $U(a, b; z)$]. One can relate the solutions at infinity and the horizon via the relation $R^\nu_0 = K^\nu R^\nu_C$, where $K^\nu$ is defined in Appendix \ref{app:MST}.

The series coefficients satisfy three-term recurrence relations of the form,
\begin{align} \label{eqn:coeff}
    \alpha^\nu_n f^\nu_{n+1} + \beta^\nu_n f^\nu_{n} + \gamma^\nu_n f^\nu_{n-1} = 0,
\end{align}
where $\alpha^\nu_n, \beta^\nu_n$, and $\gamma^\nu_n$ are given in Appendix \ref{app:MST}. The MST series converge if $\nu$ is chosen so that $f^\nu_{n}$ forms a minimal solution to Eq.~\eqref{eqn:coeff} as $|n| \rightarrow \infty$. To obtain $\nu$, one can construct the continued fractions,
\begin{subequations}
    \begin{align}
        R^\nu_n &= \frac{f^\nu_n}{f^\nu_{n-1}} = - \frac{\gamma^\nu_n}{\beta^\nu_n + \alpha^\nu_n R^\nu_{n+1}}
        \\
        L^\nu_n &= \frac{f^\nu_n}{f^\nu_{n+1}} = - \frac{\alpha^\nu_n}{\beta^\nu_n + \gamma^\nu_n L^\nu_{n-1}},
    \end{align}   
\end{subequations}
which together form an implicit equation for $\nu$,
\begin{align} \label{eqn:CFnu}
    R_n^\nu L^\nu_{n-1} = 1.
\end{align}
If $\nu$ satisfies Eq.~\eqref{eqn:CFnu} for any value of $n$, then $f^\nu_n$ is a minimal solution, because $R^\nu_n$ only converges when $f^\nu_n$ is minimal as $n \rightarrow \infty$ and $L^\nu_n$ only converges when $f^\nu_n$ is minimal as $n \rightarrow -\infty$. Note that rather than dealing with Eq.~\eqref{eqn:CFnu} directly, researchers often determine $\nu$ from the analogous equation,
\begin{align} \label{eqn:CFnu2}
    \beta^\nu_n + \alpha^\nu_n R_{n+1}^\nu +\gamma^\nu_n L^\nu_{n-1} = 0.
\end{align}
Numerical algorithms for extracting $\nu$ from Eq.~\eqref{eqn:CFnu2} can be found in \cite{FujiTago05, Thro10}. Given a value of $\nu$ that satisfies \eqref{eqn:CFnu} or \eqref{eqn:CFnu2}, the series expansions for $R^\mathrm{in}$ and $R^\mathrm{out}$ [\eqref{eqn:Rin} and \eqref{eqn:Rout}] are formally convergent on the domain $-\infty < x \leq 0$, while expansions for $R^\mathrm{up}$ and $R^\mathrm{down}$ [\eqref{eqn:Rup} and \eqref{eqn:Rdown}] converge for $ \epsilon\kappa< z \leq \infty$.

From the MST solutions, one can also construct the scattering amplitudes defined in Eqs.~\eqref{eqn:teukAsymp} and \eqref{eqn:teukAsymp2}. For example, the transmission coefficients are given by,
\begin{subequations} \label{eqn:MSTamplitudes}
    \begin{align}
        \mathcal{R}^\mathrm{in,trans} &= (2M \kappa)^{2s} e^{{i}\kappa\epsilon_+(1+\frac{2\ln \kappa}{1+\kappa})} \sum_{n=-\infty}^\infty f^\nu_n,
        \\
        \mathcal{R}^\mathrm{out,trans} &= e^{-i\kappa\epsilon_+(1+\frac{2\ln \kappa}{1+\kappa})}\sum_{n=-\infty}^{\infty} \frac{(\nu+1+i\tau)_n (\nu+1+\bar{\xi})_n}{(\nu+1-i\tau)_n (\nu+1-\bar{\xi})_n} f_{n}^{\nu},
        \\
        \mathcal{R}^\mathrm{up,trans} &= \omega^{-2s-1} A^\nu_- e^{i\epsilon (\ln \epsilon - \frac{1-\kappa}{2})},
        \\
        \mathcal{R}^\mathrm{down,trans} &= {\omega^{-1}} A^\nu_+ e^{-i\epsilon (\ln \epsilon - \frac{1-\kappa}{2})},
    \end{align}
\end{subequations}
where $A^\nu_\pm$ is defined in Appendix \ref{app:MST}. For completeness, the incidence and reflection amplitudes are also provided in Appendix \ref{app:MST}.

\section{Monodromy eigenvalues of singular points}
\label{sec:monodromy}

In general, monodromy theory focuses on the behavior of mathematical objects as they ``run around" singular points in the complex plane. For this work, we are interested in the application of monodromy theory to the solutions of second-order ODEs. This was previously studied by Castro et al.~\cite{CastETC13a, CastETC13b} in the context of scalar waves and black hole scattering, and we will ultimately connect their work on Teukolsky monodromy data to MST's renormalized angular momentum. In this section, we summarize key points from \cite{CastETC13a, CastETC13b} to provide background and establish notation.

\subsection{Background}

Following the work of \cite{CastETC13a, CastETC13b}, we consider ODEs of the form,
\begin{align} \label{eqn:ode}
	\partial_z \left[U(z) \partial_z \psi(z) \right] - V(z)\psi(z) = 0,
\end{align}
though this discussion can be extended to more generic homogeneous ODEs,
\begin{align}
    \partial_z^2 \phi(z) + f(z) \partial_z \phi(z) + g(z) \phi(z) = 0,
\end{align}
via the transformation $\phi(z) = U^{1/2}(z) e^{-\frac{1}{2}\int_z f(z') dz'} \psi(z)$. Note that the radial Teukolsky equation \eqref{eqn:teuk} already takes the form of Eq.~\eqref{eqn:ode}. Defining $\Psi = \psi$ and $\Pi = U(z)\partial_z \psi$, Eq.~\eqref{eqn:ode} can also be represented in reduced-order form by the first-order matrix equation,
\begin{align}
	\partial_z \begin{pmatrix} \Psi \\ \Pi \end{pmatrix}
	= \begin{pmatrix}
	0 & U^{-1}(z) \\
	V(z) & 0
	\end{pmatrix}
	\begin{pmatrix} \Psi \\ \Pi \end{pmatrix} \equiv
	{A}(z) \vec{\Psi},
\end{align}
where $\vec{\Psi}$ is a vector composed of $\Psi$ and $\Pi$, and the poles of $A(z)$ define the equation's singular points $z_i$. Next, let $\vec{\Psi}^{(1)}$ and $\vec{\Psi}^{(2)}$ be vectors that correspond to two independent solutions $\psi_1$ and $\psi_2$. Together these vectors form the fundamental matrix,
\begin{align} \label{eqn:matODE}
	\Phi(z) = \Big(\vec{\Psi}^{(1)} \;\; \vec{\Psi}^{(2)} \Big)
	= \begin{pmatrix}
		\psi_1 & \psi_2 \\
		U(z)\partial_z \psi_1 & U(z)\partial_z \psi_2
	\end{pmatrix}.
\end{align}
Conveniently, the determinant of this fundamental matrix is related to the constant (weighted) Wronskian of $\psi_1$ and $\psi_2$: $\mathrm{det}(\Phi) = \hat{W}(\psi_1,\psi_2) = U(z)\left(\psi_1 \partial_z \psi_2 - \psi_2 \partial_z \psi_1 \right)$. 

Next we consider the behavior of any solution $\vec{\Psi}$ as it follows a closed loop $\gamma$ (in the positive direction) in the complex domain. For the differential equations considered in this work, $A(z)$ is meromorphic (single-valued) and the operator $\partial_z - A(z)$ will always return to itself after following $\gamma$. In contrast, the fundamental matrix $\Phi$ may not return to its original value due to branch cuts of the solutions. Nonetheless, the new fundamental matrix generated by following $\gamma$, which we denote as $\Phi_\gamma$, also satisfies $[\partial_z - A(z)] \Phi_\gamma = 0$ and represents a solution to Eq.~\eqref{eqn:matODE}. Consequently, $\Phi$ and $\Phi_\gamma$ must be related by some invertible constant matrix $M_\gamma$, such that,
\begin{align}
	\Phi_\gamma(z) = \Phi(z)M_\gamma,
\end{align} 
or more explicitly,
\begin{align} \label{eqn:Mi}
	\left.\Big(\;\psi_1(z_i +e^{2\pi i}(z-z_i)) \;\; 
	\;\; \psi_2(z_i +e^{2\pi i}(z-z_i))\;\Big)
	\right\vert_{z\rightarrow z_i} =
	\left.\Big(\;\psi_1(z) \;\;  \;\; \psi_2(z)\;\Big)
	\right\vert_{z\rightarrow z_i}
	 M_i,
\end{align}
for $z_i \neq \infty$. For points at infinity, we must first perform a change of variable $\xi = 1/z$ to bring the singular point to $\xi = 0$. Circling this point is then given by $\xi \rightarrow e^{2\pi i}\xi$ or, equivalently, $z \rightarrow e^{-2\pi i} z$.

If $M_\gamma$ does not enclose a singular point, then $\gamma$ does not cross any branch cuts, and $\Phi$ will return to itself, leading to $M_\gamma = \mathbf{1}$, where $\mathbf{1}$ is the identity matrix. On the other hand, if $\gamma$ encloses one of the equation's singular points, then $M_\gamma$ will form a nontrivial transformation matrix, which we refer to as the \emph{monodromy matrix} or \emph{monodromy data} of that singular point. Crucially, the form of $M_i$ depends on the chosen basis of independent radial solutions $\psi_1$ and $\psi_2$.\footnote{Furthermore, there is residual gauge freedom in our differential equation, which will affect the values of $M_i$. In this work we work with the minimal form described in \cite{CastETC13b} (see Section 2.1 of \cite{CastETC13b} in for more details).}

A convenient property of the monodromy matrices is that, for an equation with $n$ singular points $z_1, z_2, \dots, z_n$ and $n$ monodromy matrices defined about these points $M_1, M_2, \dots, M_n$, we have
\begin{align} \label{eqn:identity}
	M_1 M_2 \cdots M_n = \mathbf{1}.
\end{align}
This identity arises from connecting the individual paths around each singular point into a single loop $\gamma'$, so that outside $\gamma'$ \emph{no} singular points are enclosed, leading to $M_1 M_2 \cdots M_n = M_{\gamma'} = \mathbf{1}$.



\subsection{Calculating monodromy data}

We now summarize relevant methods for calculating the monodromy matrices of second-order ODEs. We highlight the difference when extracting monodromy data for regular singular points (see Section \ref{sec:regular}) versus an irregular singular point (of rank one) at infinity (see Section \ref{sec:irregular}). Using these methods, we then construct the monodromy matrices associated with the singular points of the confluent hypergeometric equation (see Section \ref{sec:example}). These results will be leveraged in Sec.~\ref{sec:nuFromMono} to connect the monodromy matrices of the Teukolsky equation to the renormalized angular momentum.

\subsubsection{Regular singular points}
\label{sec:regular}

Consider solutions to Eq.~\eqref{eqn:ode}. Based on Fuchs–Frobenius theory \cite{DLMF}, one can define the behavior of these solutions near a regular singular point $z_r$ in terms of the indices $\lambda^r_{1,2}$, given by,
\begin{subequations}
    \begin{align}
        2\lambda^r_1 &= 1-f^r_0 - \sqrt{(1-f^r_0)^2-4 g^r_0},
        &
        2\lambda^r_2 &= 1-f^r_0 + \sqrt{(1-f^r_0)^2-4 g^r_0},
        \\
        f^r_0 &= \lim_{z\rightarrow z_r} \frac{(z-z_r)\partial_z U(z)}{U(z)},
        &
        g^r_0 &= \lim_{z\rightarrow z_r} -\frac{(z-z_r)^2 V(z)}{U(z)}.
    \end{align}  
\end{subequations}
If $\lambda^r_1 - \lambda^r_2 \notin \mathbb{Z}$, then there exists two independent solutions of Eq.~\eqref{eqn:ode}, $\psi^r_1(z)$ and $\psi^r_2(z)$, with series expansions, 
\begin{align} \label{eqn:regularSeries}
    \psi^r_{j}(z \rightarrow z_r) \simeq \hat{\psi}^r_{j}(z) = (z-z_r)^{\lambda^r_j} \sum_{k=0}^\infty c_{j,k} (z-z_r)^k,
\end{align}
which are convergent in a neighborhood around $z_r$. Note that $j = \{1,2\}$. Thus, we can use the series in \eqref{eqn:regularSeries} to evaluate $\psi^r_j$ after following a loop $\gamma$ around $z_r$ in the complex plane,
\begin{align} \label{eqn:psirMono}
    \psi^r_j\big(z_r +e^{2\pi i}(z-z_r)\big) = e^{2\pi i \lambda^r_j} \psi^r_j(z).
\end{align}
Combining Eq.~\eqref{eqn:psirMono} with Eq.~\eqref{eqn:Mi}, it is then straightforward to deduce the monodromy data at $z_r$,
\begin{align} \label{eqn:Mr}
    M_r \doteq M_r^{S} = \begin{pmatrix}
        e^{2\pi i \lambda^r_1} & 0, 
        \\
        0 & e^{2\pi i \lambda^r_2}
    \end{pmatrix},
\end{align}
demonstrating that $\psi^r_j$ form the basis of solutions that diagonalize $M_r$ with eigenvalues $e^{2\pi i \lambda^r_{j}}$. In the notation above, $M_r$ represents the monodromy matrix at $z_r$ in \emph{any} basis of solutions, while $M_r^{S}$ specifically refers to the form of $M_r$ in the basis of solutions with series expansions given by \eqref{eqn:regularSeries} and normalized so that $c_{j,0}=1$. A similar notation will also be used when representing the monodromy matrices of irregular singular points.

\subsubsection{Irregular singular points}
\label{sec:irregular}

Near an irregular singular point at infinity, we characterize the asymptotic behavior of solutions in terms of the characteristic roots $\mu^\infty_j$ and the indices $\lambda^\infty_j$, given by,
\begin{align}
    2\mu^\infty_1 &= - f^\infty_0 - \sqrt{(f^\infty_0)^2 - 4 g^\infty_0},
    &
    2\mu^\infty_2 &= - f^\infty_0 + \sqrt{(f^\infty_0)^2 - 4 g^\infty_0},
    &
    \lambda^\infty_j &= - \frac{f^\infty_1 \mu^\infty_j + g^\infty_1}{f^\infty_0 + 2\mu^\infty_j},
\end{align}
with,
\begin{align}
    f^\infty_i &= \lim_{z\rightarrow \infty} \partial_z^i\left[\frac{\partial_z U(z)}{U(z)} \right],
    &
    g^\infty_i &= -\lim_{z\rightarrow \infty} \partial_z^i\left[\frac{V(z)}{U(z)} \right].
\end{align}
Again, $j=\{1,2\}$ for our two independent homogeneous solutions. From these coefficients, one can define the series expansions,
\begin{align} \label{eqn:irregularSeries}
    \hat{\psi}^\infty_{j}(z) = e^{\mu^\infty_j z} (2\mu z)^{\lambda^\infty_j} \sum_{k=0}^\infty d_{j,k} z^{-k},
\end{align}
where $2\mu = \mu^\infty_2 - \mu^\infty_1$, and the series are formally non-convergent unless the series coefficients $d_{j,k}$ vanish for all $k$ above some finite $k=k_\mathrm{max}$. Provided $\mu \neq 0$, then there exists two independent solutions to Eq.~\eqref{eqn:ode}, $\psi_1^\infty$ and $\psi_2^\infty$, which are asymptotic to $\hat{\psi}^\infty_{j}(z)$ in sectors $\hat{S}_{j}$ of the complex plane \cite{DLMF, DaalOlve95},
\begin{align} \label{eqn:irregularSeriesAsymp}
    \psi^\infty_{j}(z \rightarrow \infty) &\sim \hat{\psi}^\infty_{j}(z),
    & 
    z \in \hat{S}_{j}.
\end{align}
These wedges $\hat{S}_{j}$ are defined by the (anti-)Stokes lines, such that,
\begin{align} \label{eqn:Sj}
	\hat{S}_j = \left\{ 
	z : \left(j-\frac{5}{2}\right)\pi +\delta \leq \mathrm{ph}(2\mu z)
	\leq \left(j+\frac{1}{2}\right)\pi -\delta
	\right\},
\end{align}
where $\mathrm{ph}(z)$ is the phase of $z$, and $0<\delta \ll 1$. (See Fig.~1.1 in Ref.~\cite{DaalOlve95} for a visualization of the related subsectors $S_j$, which are connected to those in Eq.~\eqref{eqn:Sj} by $\hat{S}_j=S_{j-2}\cup S_{j-1} \cup S_{j}$.) To analytically continue solutions around $z=\infty$, one can make use of the connection formulae\footnote{See Appendix \ref{app:connection} for alternate forms.} \cite{DLMF, DaalOlve94, DaalOlve95},
\begin{subequations} \label{eqn:connectionFormulae}
    \begin{align} \label{eqn:connection1}
        \psi^\infty_1(z) &= e^{2\pi i \lambda^\infty_1} \psi^\infty_1(e^{-2\pi i}z) - C_1\psi^\infty_2(z),
        \\ \label{eqn:connection2}
        \psi^\infty_2(z) &= e^{-2\pi i \lambda^\infty_2} \psi^\infty_2(e^{2\pi i}z) + C_2\psi^\infty_1(z),
    \end{align}
\end{subequations}
where $C_1$ and $C_2$ are the well-known \emph{Stokes multipliers}\footnote{See Refs.~\cite{DLMF, DaalOlve94, DaalOlve95} for further discussion on the role of Stokes multipliers in second-order ODEs. Note that we use notation similar to that of Refs.~\cite{DLMF, DaalOlve94} but with $C_k$ replaced by $(-1)^k C_k$. This differs from the notation in Ref.~\cite{DaalOlve95}, as described in footnote 4 of \cite{DaalOlve95}. Therefore, $C_1$ and $C_2$ in this work are equivalent to $C_1$ and $C_0$, respectively, in Ref.~\cite{DaalOlve95}.}, which can be determined from the limits,
\begin{align} \label{eqn:StokeMultipliersFromLimits}
    C_1 &= - \frac{2\pi i (2\mu)^{2\lambda} e^{-2\pi i \lambda}}{d_{2,0}} \lim_{n\rightarrow\infty} \frac{(-2\mu)^n d_{1,n}}{\Gamma(n+2\lambda)},
    &
    C_2 &= - \frac{2\pi i (2\mu)^{-2\lambda}}{d_{1,0}} \lim_{n\rightarrow\infty} \frac{(2\mu)^n d_{2,n}}{\Gamma(n-2\lambda)},
\end{align}
with $2\lambda = \lambda_2^\infty - \lambda_1^\infty$ and the coefficients $d_{j,k}$ defined by the series expansion in \eqref{eqn:irregularSeriesAsymp}.
Consequently, the terms on the right-hand side of Eq.~\eqref{eqn:connection1} are asymptotic to the series \eqref{eqn:irregularSeries} for $z \in \hat{S}_{2} \cap \hat{S}_{3}$, while the terms in Eq.~\eqref{eqn:connection2} are asymptotic to \eqref{eqn:irregularSeries} for $z \in \hat{S}_{0} \cap \hat{S}_{1}$.

We then combine these results to evaluate the solutions $\psi^\infty_j$ after circling $z=\infty$, 
\begin{subequations} \label{eqn:infinityRelationR1R2}
    \begin{align}
    	\psi^\infty_{1} (e^{-2\pi i}z)
    	&= e^{-2\pi i \lambda_1^\infty} \left[{\psi}^\infty_1(z) + C_1{\psi}^\infty_2(z) \right],
    	\\
    	\psi^\infty_{2} (e^{-2\pi i}z)
    	&= e^{-2\pi i \lambda_2^\infty} \left[ C_2 e^{4\pi i \lambda} {\psi}^\infty_1(z) + (1+C_1 C_2 e^{4\pi i \lambda}) {\psi}^\infty_2(z)
    	\right].
    \end{align}
\end{subequations}
From \eqref{eqn:infinityRelationR1R2}, the monodromy matrix $M_\infty$ then takes the form (in the basis of solutions $\psi^\infty_{1,2}$),
\begin{align} \label{eqn:Minf}
	M_\infty \doteq M^{S}_\infty = e^{-\pi i(\lambda_1^\infty +\lambda_2^\infty)} \hat{M}^{S}_\infty = e^{-\pi i(\lambda_1^\infty +\lambda_2^\infty)}\begin{pmatrix}
	e^{2\pi i \lambda} & e^{2\pi i \lambda} C_1
	\\
	e^{2\pi i \lambda}C_2 & e^{-2\pi i \lambda}\left[1 + C_1 C_2 e^{4\pi i \lambda} \right]
	\end{pmatrix},
\end{align}
where for convenience we have defined the normalized matrix $\hat{M}^{S}_\infty$ with determinant $\mathrm{det}(\hat{M}^{S}_\infty) = 1$. Similar to $M_r$ and $M^\mathrm{S}_r$ in Sec.~\ref{sec:regular}, ${M}_\infty$ represents the monodromy matrix at infinity for any basis of solutions, while ${M}^{S}_\infty$ is the specific form of the monodromy matrix for the solutions $\hat{\psi}^\infty_{1,2}$, which are asymptotic to the series expansions \eqref{eqn:irregularSeries} with normalizations $d_{j,0}=1$. 

Inspecting Eq.~\eqref{eqn:Minf}, we immediately observe that $\psi^\infty_j$ does not diagonalize $M_\infty$ despite $\psi^\infty_j$ being the natural basis for describing the asymptotic behavior of solutions. However, one can still construct solutions that diagonalize $M_\infty$, which we refer to as $\psi^{\infty}_{D,j}$, by relating them to $\psi^\infty_j$ using the eigenvectors and eigenvalues of Eq.~\eqref{eqn:Minf}. To solve for the monodromy eigenvalues, we follow Refs.~\cite{CastETC13a, CastETC13b} and consider that, in the basis of $\psi^{\infty}_{D,j}$, the monodromy matrix takes the form,
\begin{align}
    M^{D}_\infty = e^{-\pi i (\nu_1^\infty+ \nu_2^\infty)} \hat{M}^{D}_\infty = e^{-\pi i (\nu_1^\infty+ \nu_2^\infty)}\begin{pmatrix}
	e^{2\pi i \nu_\infty} & 0
	\\
	0 & e^{-2\pi i \nu_\infty}
	\end{pmatrix},
\end{align}
where $e^{-2\pi i \nu^\infty_j}$ are the monodromy eigenvalues of $\psi^{\infty}_{D,j}$, $2\nu_\infty = \nu_2^\infty - \nu_1^\infty$, and again we define the normalized matrix $\hat{M}^D_\infty$ with unit determinant. Because $M^D_\infty$ is equivalent to $M^S_\infty$ up to a change of basis, we equate the determinants and traces of both matrices, leading to,
\begin{align} \label{eqn:monodromyEigenvalues}
    e^{-2\pi i (\nu_1^\infty+ \nu_2^\infty)} &= e^{-2\pi i(\lambda_1^\infty +\lambda_2^\infty)},
    &
    2\cos 2\pi\nu_\infty &= 2\cos 2\pi\lambda + e^{2\pi i\lambda} C_1 C_2,
\end{align}
from which one can calculate $\nu_j^\infty$ given the combination $C_1 C_2$. Note that Eq.~\eqref{eqn:monodromyEigenvalues} is particularly useful for extracting the eigenvalues, because the combination $C_1 C_2$ does not depend on the overall normalizations of $\psi^\infty_1$ and $\psi^\infty_2$ even though $C_1$ and $C_2$ do, individually. [This is evident from Eq.~\eqref{eqn:StokeMultipliersFromLimits}.] 

\subsubsection{Monodromy data of confluent hypergeometric functions}
\label{sec:example}

Now we apply these methods to extract the monodromy data of the confluent hypergeometric equation,\footnote{Instead of bringing Eq.~\eqref{eqn:cfhODE} into the form of Eq.~\eqref{eqn:ode}, for example via the transformation $w(z) = e^{z/2} z^{-b/2} W(z)$, we simplify our calculations by directly applying the methods from Sec.~\ref{sec:regular} and Eq.~\ref{sec:irregular} to the solutions $w(z)$. One can verify that performing this analysis for $W(z)$ or $w(z)$ leads to consistent results for the monodromy data.}
\begin{align} \label{eqn:cfhODE}
    z \frac{d^2 w}{dz^2} + (b - z) \frac{d w}{dz} - a w = 0,
\end{align}
which possesses a regular singular point at $z=0$ and irregular point at $z=\infty$.
Standard solutions to Eq.~\eqref{eqn:cfhODE} include the regular and irregular confluent hypergeometric functions, $M(a, b; z)$ and $U(a, b; z)$, first introduced in Sec.~\ref{sec:MST}. As their names suggest, $M(a, b; z)$ is regular at $z=0$ and is represented by the series solution,
\begin{align} \label{eqn:CHFM}
    M(a, b, z) = \sum_{k=0}^\infty \frac{(a)_k}{(b)_k}\frac{z^k}{k!},
\end{align}
which is entire in $z\in \mathbb{C}$, while $U(a, b; z)$ is associated with the irregular singular point and is defined by its asymptotic behavior as $z \rightarrow \infty$,
\begin{align} \label{eqn:CHFU}
    U(a, b, z) \sim z^{-a} \sum_{k=0}^\infty (-1)^k \frac{(a)_k (a-b+1)_k}{k!} z^{-k}.
\end{align}

First we analyze the monodromy matrix at $z=0$. Solutions near the regular singular point are defined by the indices $\lambda_1^{0}=0$ and $\lambda_2^{0}=1-b$, with series representations,
\begin{align} \label{eqn:chfRegSeries}
    {w}^0_1 (z\rightarrow 0) \simeq \hat{w}^0_1 (z) &= \sum_{k=0}^\infty c_{1,k} z^k,
    &
    {w}^0_2 (z\rightarrow 0) \simeq \hat{w}^0_2 (z) &= z^{1-b}\sum_{k=0}^\infty c_{2,k} z^k.
\end{align}
In the case of the hypergeometric functions, the series coefficients take the compact forms,
\begin{align}
    c_{1,k} &= \frac{(a)_k}{(b)_k k!},
    &
    c_{2,k} &= \frac{(a-b+1)_k}{(2-b)_k k!},
\end{align}
from which one can identify $w_1^0(z) = \hat{w}_1^0(z) = M(a, b; z)$ and $w_2^0(z) = \hat{w}_2^0(z) = z^{1-b} M(a-b+1, 2-b; z)$. From Eq.~\eqref{eqn:Mr}, it is straightforward to assemble the monodromy matrix at $z=0$,
\begin{align} \label{eqn:M0}
    M_0^S = \begin{pmatrix}
        1 & 0, 
        \\
        0 & e^{-2\pi i b}
    \end{pmatrix}.
\end{align}

To construct the monodromy matrix of the irregular singular point $M_\infty$, we consider that the solutions near $z=\infty$ are defined by the characteristic roots $\mu^\infty_1 = 0$ and $\mu^\infty_2 = 1$ and the indices  $\lambda^\infty_1 = -a$ and $\lambda^\infty_2 = a-b$, with asymptotic solutions,
\begin{align} \label{eqn:chfIrregSeries}
    {w}^\infty_1 (z\rightarrow \infty) \sim \hat{w}^\infty_1 (z) &= z^{-a}\sum_{k=0}^\infty d_{1,k} z^{-k},
    &
    {w}^\infty_2 (z\rightarrow \infty) \sim \hat{w}^\infty_2 (z) &= e^{z}z^{a-b}\sum_{k=0}^\infty d_{2,k} z^{-k}.
\end{align}
Once again, the series coefficients have the compact forms,
\begin{align}
    d_{1,k} &= (-1)^k \frac{(a)_k (a-b+1)_k}{k!},
    &
    d_{2,k} &= \frac{(b-a)_k (1-a)_k}{k!},
\end{align}
leading to $w_1^\infty(z) = U(a, b; z)$ and $w_2^\infty(z) = e^{\pm i \pi(a-b)} e^z U(b-a, b, e^{\mp i \pi} z)$.

Because we have analytic expressions for the asymptotic coefficients of $w_1^\infty$ and $w_2^\infty$, we can directly evaluate the Stokes multipliers using \eqref{eqn:StokeMultipliersFromLimits}, leading to
\begin{align} \label{eqn:C1andC2chf}
    C_1 &= -\frac{2\pi i e^{-\pi i(2a-b)}}{\Gamma(a)\Gamma(a-b+1)},
    &
    C_2 &= -\frac{2\pi i}{\Gamma(b-a)\Gamma(1-a)}
\end{align}
and
\begin{align} \label{eqn:C1C2chf}
    C_1 C_2 e^{2\pi i \lambda} &= - \frac{4\pi^2}{\Gamma(a)\Gamma(1-a)\Gamma(b-a)\Gamma(1-b+a)} = -4 \sin\pi a\sin\pi(b-a).
\end{align}
Combining Eq.~\eqref{eqn:C1andC2chf} and Eq.~\eqref{eqn:C1C2chf} with the fact that $2\lambda = 2a-b$ and $\lambda_1^\infty + \lambda_2^\infty = -b$, it is straightforward to read-off the monodromy matrix via Eq.~\eqref{eqn:Minf}. As expected, we observe that $U(a,b,z)$ and $e^z U(b-a,b,-z)$ do not diagonalize $M_\infty$, as discussed in Sec.~\ref{sec:irregular}.

Similarly, we can calculate the eigenvalues with Eq.~\eqref{eqn:monodromyEigenvalues}, yielding
\begin{align}
    e^{-2\pi i(\nu_1 + \nu_2)} &= e^{-2\pi i(\lambda^\infty_1+\lambda^\infty_2)},
    &
    \cos2\pi\nu_\infty &= \cos\pi b,
\end{align}
or
\begin{align}
    \nu_1 + \nu_2 &= -b + n',
    &
    \nu_2^\infty - \nu_1^\infty &= \pm ( b + 2 k'),
\end{align}
for arbitrary integers $k'$ and $n'$. Choosing $k'=n'=0$, we have $\nu^\infty_1 = 0$ and $\nu^\infty_2 = -b$, leading to the diagonalized monodromy matrix,
\begin{align} \label{eqn:MchfInfDiagonal}
    M^D_\infty = \begin{pmatrix}
        1 & 0
        \\
        0 & e^{2\pi i b}
    \end{pmatrix}.
\end{align}
We can also derive \eqref{eqn:MchfInfDiagonal} using the identity Eq.~\eqref{eqn:identity}. When both monodromy matrices share a common basis, ${M}_0 {M}_\infty = \mathbf{1}$ or ${M}_\infty = {M}_0^{-1}$. Comparing Eqs.~\eqref{eqn:MchfInfDiagonal} and \eqref{eqn:M0}, we find that $M^D_\infty = (M^S_0)^{-1}$. Thus they share the same basis, and the solutions $M(a, b, z)$ and $z^{1-b} M(a-b+1,2-b,z)$ diagonalize both matrices.\footnote{Furthermore, because the series representation \eqref{eqn:CHFM} of $M(a,b;z)$ is entire in $z$, we can use \eqref{eqn:CHFM} to directly evaluate $M(a,b; e^{-2\pi i}z)$. This also makes it apparent that $w_1^0(z)$ and $w_2^0(z)$ diagonalize the monodromy matrices of both singular points.}

\section{Monodromy eigenvalues of the Teukolsky equation and their relation to renormalized angular momentum}
\label{sec:nuFromMono}

We now apply the monodromy methods of Sec.~\ref{sec:monodromy} to the radial Teukolsky equation \eqref{eqn:teuk}. First we construct the monodromy matrix at $r=r_+$, which we denote by $M_\mathcal{H}$. In the basis of $R^\mathrm{in}$ and $R^\mathrm{out}$, the monodromy matrix is explicitly defined by the transformation,
\begin{align*}
    \left.\Big(\;R^\mathrm{in}[r_+ +e^{2\pi i}(r-r_+)] \;\;, 
	\;\; R^\mathrm{out}[r_+ +e^{2\pi i}(r-r_+)]\;\Big)
	\right\vert_{r\rightarrow r_+} 
    =
	\left.\Big(\;R^\mathrm{in}(r) \;\; , \;\; R^\mathrm{out}(r)\;\Big)\right\vert_{r\rightarrow r_+} M^S_\mathcal{H}.
\end{align*}
Here we continue the notation established in Sec.~\ref{sec:monodromy} and use $M^S_\mathcal{H}$ to represent the form of $M_\mathcal{H}$ in this natural choice of basis. From the expansions in \eqref{eqn:horizonSeries}, we observe that these horizon solutions possess the singular indices $\lambda^\mathcal{H}_\mathrm{out} = i\epsilon_+$ and $\lambda^\mathcal{H}_\mathrm{in} = -s-i\epsilon_+$. Thus $e^{\pm 2\pi \epsilon_+}$ are the monodromy eigenvalues at the horizon, leading to the representation $M^S_\mathcal{H} = \mathrm{diag}(e^{2\pi \epsilon_+}, e^{-2\pi \epsilon_+}$).

Next we examine the monodromy matrix at infinity, which we refer to as $M_\mathcal{I}$. From the expansions in \eqref{eqn:cheIrregular}, we can read-off the characteristic roots $2\mu^\mathcal{I}_\mathrm{up} = 0$ and $2\mu^\mathcal{I}_\mathrm{down} = -2i\kappa \epsilon$ and the indices $\lambda^\mathcal{I}_\mathrm{up} = -1-2s$ and $\lambda^\mathcal{I}_\mathrm{down} = -1-2i\epsilon$ for the infinity solutions $R^\mathrm{up}(z)$ and $R^\mathrm{down}(z)$. This leads to $\mu = -i\kappa\epsilon$ and $\lambda = \xi$.\footnote{Alternatively, from the $r$-coordinate expansions in \eqref{eqn:infinitySeries}, we have roots $2\mu^\mathcal{I}_\mathrm{up} = i\epsilon$ and $2\mu^\mathcal{I}_\mathrm{down} = -i\epsilon$ and indices $\lambda^\mathcal{I}_\mathrm{up} = -1-2s+i\epsilon$ and $\lambda^\mathcal{I}_\mathrm{down} = -1-i\epsilon$, leading to $2\mu=-i\epsilon$ and $2\lambda=2\xi$.} By calculating the associated Stokes multipliers $C_1$ and $C_2$, one can also construct $M^S_\mathcal{I}$ via Eq.~\eqref{eqn:Minf}, leading to,
\begin{align} \label{eqn:MinfMST}
	\left.\Big(\;R^\mathrm{up}(e^{-2\pi i}z) \;\; ,
	\;\; R^\mathrm{down}(e^{-2\pi i}z)\;\Big)
	\right\vert_{z\rightarrow \infty} =
	\left.\Big(\;R^\mathrm{up}(z) \;\; ,  \;\; R^\mathrm{down}(z)\;\Big)\right\vert_{z\rightarrow \infty}
	 M^S_\mathcal{I},
\end{align}
where we now use the MST radial coordinate defined in Eq.~\eqref{eqn:MSTcoord} to study the infinity-side solutions. 

Because the MST expansions are composed of analytic functions, we can also use them to directly evaluate $R^\mathrm{up}(e^{-2\pi i} z)$ and $R^\mathrm{down}(e^{-2\pi i}z)$ and extract the monodromy eigenvalues at infinity. To simplify this calculation, we first investigate the monodromy data of $R^\nu_C(z)$ and $R^{-\nu-1}_C(z)$. In Sec.~\ref{sec:example}, we demonstrated that $M(a,b;z)$---rather than $U(a,b;z)$---diagonalizes the monodromy matrix at infinity for the confluent hypergeometric equation. Thus, we might expect that $R^\nu_C(z)$ and $R^{-\nu-1}_C(z)$---which depend on $M(a,b;z)$ [see Eq.~\eqref{eqn:RnuC}]---form a natural basis for examining the monodromy data of the Teukolsky equation. Because the series \eqref{eqn:RnuC} is analytic and convergent at $z=\infty$, it is straightforward to evaluate $R^\nu_C(e^{-2\pi i}z)$ as $z\rightarrow \infty$,
\begin{align} \label{eqn:Rnu2piI}
    R^\nu_C(e^{-2\pi i}z) &= e^{-2\pi i\nu} R^\nu_C(z),
    &
    R^{-\nu-1}_C(e^{-2\pi i}z) &= e^{2\pi i\nu} R^{-\nu-1}_C(z),
    & 
    (z &\rightarrow \infty).
\end{align}
Immediately, we see that $R^\nu_C$ and $R^{-\nu-1}_C$ do in fact form a basis the diagonalizes $M_\mathcal{I}$, and $e^{\mp 2\pi i\nu}$ are their monodromy eigenvalues. Thus, (up to some arbitrary integer) the renormalized angular momentum $\nu$ \emph{is the (logarithm of the) monodromy eigenvalue at infinity} (as well as its reflected value $-\nu-1$), i.e., $\nu = \pm \nu_\infty + N_\infty$ for $N_\infty \in \mathbb{Z}$.

Combining Eq.~\eqref{eqn:Rnu2piI} with \eqref{eqn:RupToRnu}, we can extract $M^S_\mathcal{I}$ from Eq.~\eqref{eqn:MinfMST},
\begin{align}
    M_\mathcal{I}^S = \begin{pmatrix}
        e^{-2\pi i \xi} & 2i e^{-i\pi(\nu + \xi)} \mathcal{B}^\mathrm{trans} \sin\pi(\nu-\xi)
        \\
        2i e^{i\pi(\nu - \xi)} \left(\mathcal{B}^\mathrm{trans}\right)^{-1} \sin\pi(\nu+\xi) & 2\cos 2\pi\nu - e^{-2i\pi\xi}
    \end{pmatrix},
\end{align}
where $\mathcal{B}^\mathrm{trans} = \mathcal{R}^\mathrm{down,trans}/\mathcal{R}^\mathrm{up,trans}$. This leads to the Stokes multipliers,
\begin{align}
    C_1 &= 2i e^{-i\pi(\nu+3\xi)} \mathcal{B}^\mathrm{trans} \sin\pi(\nu - \xi),
    &
    C_2 &= 2i e^{i\pi(\nu-3\xi)} \left(\mathcal{B}^\mathrm{trans}\right)^{-1} \sin\pi(\nu + \xi).
\end{align}
Furthermore, from these solutions we can verify that $\mathrm{det}(M^S_\mathcal{I}) = 1$ and $\mathrm{Tr}(M^S_\mathcal{I}) = 2\cos2\pi\nu = 2\cos2\pi\nu_\infty$, as expected.


\section{Numerical methods for calculating the monodromy eigenvalues of the radial Teukolsky equation}
\label{sec:numerical}

We now highlight numerical methods for extracting $\nu$ via the monodromy eigenvalue equation \eqref{eqn:monodromyEigenvalues}. In Sec.~\ref{sec:C1C2}, we outline a numerical procedure for calculating the combination $C_1 C_2 e^{2\pi i \lambda}$ based on the work of Daalhuis and Olver \cite{DaalOlve95}. In Sec.~\ref{sec:results} we provide numerical calculations of $\nu$, which are obtained with these monodromy methods, and we compare our numerical calculations against those reported in Ref.~\cite{CastETC13b}. Then in Sec.~\ref{sec:stability}, we discuss the numerical stability of Eq.~\eqref{eqn:monodromyEigenvalues} and highlight regions of parameter space where \eqref{eqn:monodromyEigenvalues} experiences catastrophic cancellations. We also propose methods for mitigating these numerical issues.

\subsection{Calculating Stokes multipliers for the Teukolsky equation}
\label{sec:C1C2}

Unlike the case of the confluent hypergeometric functions in Sec.~\ref{sec:example}, we cannot directly evaluate $C_1$ and $C_2$ from Eq.~\eqref{eqn:StokeMultipliersFromLimits}, because we do not have analytic expressions for the series coefficients that define the asymptotic behavior of the MST solutions $R^\mathrm{up}(z)$ and $R^\mathrm{down}(z)$. Instead, we approximate the Stokes multipliers using the results derived in Ref.~\cite{DaalOlve95}, leading to the expressions,
\begin{subequations} \label{eqn:C1C2Approx}
    \begin{align}
    C_1 &= -2\pi i \hat{b}_{1,s} (2\mu)^{2\lambda} e^{-2\pi i \lambda}\left\{\sum_{n=0}^{m-1} (-2\mu)^{n-s} \hat{b}_{2,n} \Gamma(s+2\lambda-n) \right\}^{-1} + O(s^{-m}),
    \\
    C_2 &= -2\pi i \hat{b}_{2,s} (2\mu)^{-2\lambda} \left\{\sum_{n=0}^{m-1} (2\mu)^{n-s} \hat{b}_{1,n} \Gamma(s-2\lambda-n) \right\}^{-1} + O(s^{-m}),
\end{align}
\end{subequations}
for fixed integers $s$ and $m$.\footnote{Note that as $s\rightarrow \infty$, the $n=0$ term dominates each sum, and Eq.~\eqref{eqn:C1C2Approx} reduces to Eq.~\eqref{eqn:StokeMultipliersFromLimits}. In fact, Ref.~\cite{DaalOlve95} uses Eq.~\eqref{eqn:C1C2Approx} to derive \eqref{eqn:StokeMultipliersFromLimits}, rather than first defining the Stokes multipliers in terms of these asymptotic limits, as we have done in this work.} Here, $\hat{b}_{j,n}$ represent the asymptotic series coefficients for $R^\mathrm{up}(z)$ and $R^\mathrm{down}(z)$.

To simplify the calculation of $C_1$ and $C_2$, we make two adjustments to Eq.~\eqref{eqn:C1C2Approx}. First, rather than calculating the Stokes multipliers for $R^\mathrm{up}(z)$ and $R^\mathrm{down}(z)$, we instead solve for those associated with the confluent Heun solutions $w^\mathrm{up}(\hat{z})$ and $w^\mathrm{down}(\hat{z})$. Recall that these functions are related to the MST solutions via \eqref{eqn:cheTransform}. The advantage of the confluent Heun solutions is that the coefficients of their asymptotic expansions [see Eq.~\eqref{eqn:cheIrregular}] satisfy simple three-term recurrence relations given in \eqref{eqn:cheRecurrence}. Thus they are much more efficient to numerically calculate. Furthermore, as shown in Appendix \ref{app:connection}, the transformation \eqref{eqn:cheTransform} preserves the values of $C_1$ and $C_2$ (provided both sets of solutions are normalized to the same values at the boundaries). Therefore, by calculating $C_1$ and $C_2$ for $w^\mathrm{up}(\hat{z})$ and $w^\mathrm{down}(\hat{z})$, we also obtain the Stokes multipliers for the Teukolsky solutions.

Second, we introduce the auxiliary coefficients,
\begin{align} \label{eqn:cs}
    c^{s}_{1,n} &= (2\mu)^{n-s}\hat{b}_{1,n}(-2\lambda)_{s-n},
    &
    c^{s}_{2,n} &= (-2\mu)^{n-s} \hat{b}_{2,n}(2\lambda)_{s-n},
\end{align}
which, when combined with Eq.~\eqref{eqn:C1C2Approx}, leads to,
\begin{align}
    C_1 &\simeq  -\frac{2\pi i (2\mu)^{2\lambda}
	e^{-2\pi i\lambda} \, c^s_{1,s}}{\Gamma(2\lambda)} \left\{\sum_{n=0}^{m_1-1}
	c^s_{2,n} \right\}^{-1},
    &
	C_2 & \simeq -\frac{{2\pi i (2\mu)^{-2\lambda}}  c^s_{2,s}}{\Gamma(-2\lambda)}
	\left\{ \sum_{k=0}^{m_2-1}
	c^s_{1,k} \right\}^{-1}.
\end{align}
This provides a compact expression for the combination $C_1 C_2 e^{2\pi i \lambda}$ in Eq.~\eqref{eqn:monodromyEigenvalues},
\begin{align} \label{eqn:C1C2compact}
  C_1 C_2 e^{2\pi i \lambda} = 8\pi \lambda \sin 2\pi\lambda\, c^s_{1,s} c^s_{2,s}
 	\left\{\sum_{n=0}^{m_1-1} \sum_{k=0}^{m_2-1} c^s_{1,k}c^s_{2,n}
 	\right\}^{-1}.
\end{align}
Note that the coefficients $c^s_{j,n}$ also satisfy three-term recurrence relations, but they are numerically unstable. Instead one can alternate between the recurrence equation for $\hat{b}_{j,n}$ given in \eqref{eqn:cheRecurrence} and the stepping relation,
\begin{align}
    c^{s+1}_{j,n} = (-1)^{j+1}\left(\frac{(-1)^{j} 2\mu+s-n}{2\lambda}\right) c^s_{j,n},
\end{align}
to simultaneously construct $\hat{b}_{j,n}$ and $c^{s}_{j,n}$ for $0 \leq n \leq s$.
For large values of $s$, it is also advantageous to normalize the weighted coefficients $c^{s}_{j,n}$ at each step in the recurrence so that $c^{s}_{j,s} = 1$. This avoids numerical overflow issues when taking the ratio $c^s_{j,s}/c^s_{j,0}$ due to $c^s_{j,s}$ and $c^s_{j,0}$ both growing as $\sim \Gamma(s)$. One can then vary $m$ and $s$ until \eqref{eqn:C1C2compact} converges to a numerically satisfactory value. A simple approach is to set $m = \mathrm{ceil}[s/2]$---where $\mathrm{ceil}[x]$ is the closest integer greater than $x$---and increase $s$ until the value of $C_1 C_2 e^{2\pi i \lambda}$ does not change within some numerical tolerance. Alternatively, one can choose $m$ so that the sum truncates at the coefficient $c^{s}_{j,n}$ with the smallest magnitude (for fixed $s$).

\subsection{Extracting the renormalized angular momentum}
\label{sec:results}

We present numerical results for the renormalized angular momentum $\nu$ based on the computation of the monodromy parameter $\nu_\infty$ in Eq.~\eqref{eqn:monodromyEigenvalues}. Because $\nu$ is not uniquely defined by the MST constraint equations \eqref{eqn:CFnu} and \eqref{eqn:CFnu2}\footnote{One has the freedom to shift the value of $\nu$ by an integer or flip its sign, and it will still lead to convergent MST series solutions.} and due to branch cuts in $\arccos z$, there is residual freedom in relating $\nu$ and $\nu_\infty$. In this work, we relate the two parameters via,
\begin{align} \label{eqn:nuToNuInf}
    \nu &= l - \Delta\nu,
    &
    \Delta\nu &= \arccos(\cos2\pi\nu_\infty),
\end{align}
where $\arccos z$ takes its principal values as defined in Ref.~\cite{DLMF}. Through this choice, \eqref{eqn:nuToNuInf} is consistent with low-frequency expansions of $\nu$ reported in the post-Newtonian literature (e.g, \cite{SasaTago03, KavaOtteWard16, Munn20}).

First, we reproduce the monodromy eigenvalues for the various quasinormal mode frequencies reported in Tables B1 and B2 of Ref.~\cite{CastETC13b}.\footnote{Ref.~\cite{CastETC13b} calculates the eigenvalue $\alpha_\mathrm{irr}$, which is related to our monodromy eigenvalue by $\nu_\infty = -i\alpha_\mathrm{irr}$. Additionally, their spheroidal eigenvalue $K_l$ is related to the eigenvalue in Eq.~\eqref{eqn:teuk} via $\lambda^T_{slm\omega} = K_l + a^2\omega^2 - 2m a \omega$.} In Table \ref{tab:QNM-comp}, we report the monodromy eigenvalue, $\nu_\infty^{(1)}$, based on the frequencies $\omega_{lmn}^{\mathrm{QNM}(1)}$ used in Ref.~\cite{CastETC13b}. We also report the relative difference between $\nu_\infty^{(1)}$ and the values given in Ref.~\cite{CastETC13b}. Because these quasinormal mode frequencies are less accurate in the near-extremal limit $(a\gtrsim 0.999)$, in Table \ref{tab:QNM-comp-2} we compute a second value $\nu_\infty^{(2)}$ based on the quasinormal frequencies $\omega_{lmn}^{\mathrm{QNM}(2)}$ produced by the Python package \texttt{qnm} \cite{Stei19} as a comparison.

{\renewcommand{\arraystretch}{1.25}
\begin{table*}[]
    \centering
    \caption{Monodromy eigenvalues for the quasinormal mode frequencies reported in Tables B1 and B2 of Ref.~\cite{CastETC13b}. We label overtones $n$ using the conventions of \texttt{qnm} \cite{Stei19}. The frequencies $M\omega_{lmn}^{\mathrm{QNM}(1)}$ are the quasinormal mode frequencies published in Ref.~\cite{CastETC13b}. The monodromy eigenvalues produced by these frequencies are given by $\nu^{(1)}_\infty$. In the last column we give the relative difference $|1-\cos2\pi\nu^{(1)}_\infty/\cosh2\pi\alpha_\mathrm{irr}|$, where $\alpha_\mathrm{irr}$ is the monodromy eigenvalue reported in Ref.~\cite{CastETC13b}. The values of $M\omega_{lmn}^{\mathrm{QNM}(1)}$ and $\nu^{(1)}_\infty$ are truncated below $10^{-6}$ for brevity.}
    \setlength\tabcolsep{0pt}
    \begin{tabular*}{\linewidth}{@{\extracolsep{\fill}} c c c c c c c}
        \hline
        $a/M$ & $l$ & $m$ & $n$ & $M\omega_{lmn}^{\mathrm{QNM}(1)}$ & $\nu_\infty^{(1)}$ & rel.~diff \cite{CastETC13b}
        \\
        \hline
        $0.0$ & $0$ & $0$ & $0$ & $0.110455 - 0.104896i$ & $-0.004894 - 0.106880i$ & \num{1.1e-06} \\
        $0.0$ & $0$ & $0$ & $1$ & $0.086117 - 0.348052i$ & $-0.395024 - 0.184325i$ & \num{2.3e-06} \\
        $0.0$ & $0$ & $0$ & $2$ & $0.075742 - 0.600080i$ & $-0.141796 + 0.196084i$ & \num{1.3e-02} \\
        $0.0$ & $0$ & $0$ & $2$ & $0.075742 - 0.601080i$ & $-0.139820 + 0.196497i$ & \num{2.1e-05} \\
        $0.2$ & $0$ & $0$ & $0$ & $0.110768 - 0.104512i$ & $-0.004188 - 0.106923i$ & \num{3.1e-02} \\
        $0.4$ & $0$ & $0$ & $0$ & $0.111699 - 0.103253i$ & $-0.001950 - 0.106939i$ & \num{1.4e-06} \\
        $0.6$ & $0$ & $0$ & $0$ & $0.113171 - 0.100698i$ & $-0.002224 + 0.106438i$ & \num{1.8e-06} \\
        $0.8$ & $0$ & $0$ & $0$ & $0.114537 - 0.095701i$ & $-0.008975 + 0.103590i$ & \num{1.1e-07} \\
        $0.96$ & $0$ & $0$ & $0$ & $0.111452 - 0.089387i$ & $-0.012920 + 0.094805i$ & \num{4.0e-07} \\
        $0.98$ & $0$ & $0$ & $0$ & $0.110616 - 0.089481i$ & $-0.012110 + 0.094050i$ & \num{8.4e-07} \\
        $0.99$ & $0$ & $0$ & $0$ & $0.110447 - 0.089499i$ & $-0.011954 + 0.093895i$ & \num{8.1e-07} \\
        $0.999$ & $0$ & $0$ & $0$ & $0.110384 - 0.089398i$ & $-0.012009 + 0.093741i$ & \num{1.0e-06} \\
        $0.9999$ & $0$ & $0$ & $0$ & $0.109263 - 0.090699i$ & $-0.009743 + 0.093760i$ & \num{1.0e-06} \\
        $0.9999$ & $2$ & $2$ & $0$ & $0.993235 - 0.003525i$ & $\phantom{-}1.501539 + 0.944336i$ & \num{1.2e-05} \\
        $0.9999$ & $2$ & $2$ & $1$ & $0.993220 - 0.010597i$ & $\phantom{-}1.504795 + 0.946068i$ & \num{4.9e-06} \\
        $0.9999$ & $2$ & $2$ & $2$ & $0.993175 - 0.017657i$ & $\phantom{-}1.508539 + 0.949417i$ & \num{1.0e-05}
        \\
        \hline
    \end{tabular*}
    \label{tab:QNM-comp}
\end{table*}
}

{\renewcommand{\arraystretch}{1.25}
\begin{table*}[]
    \centering
    \caption{Monodromy eigenvalues for the quasinormal mode frequencies reported in Tables B2 of Ref.~\cite{CastETC13b} for $a/M = 0.9999$. We label overtones $n$ using the conventions of \texttt{qnm} \cite{Stei19}. The frequencies $M\omega_{lmn}^{\mathrm{QNM}(2)}$ refer to the quasinormal mode frequencies calculated by \texttt{qnm}. The monodromy eigenvalues produced by these frequencies are given by $\nu^{(2)}_\infty$.}
    \setlength\tabcolsep{0pt}
    \begin{tabular*}{\linewidth}{@{\extracolsep{\fill}} c c c c c c}
        \hline
        $a/M$ & $l$ & $m$ & $n$ & $M\omega_{lmn}^{\mathrm{QNM}(2)}$ & $\nu_\infty^{(2)}$
        \\
        \hline
        $0.9999$ & $0$ & $0$ & $0$ & $0.110244 - 0.178865i$ & $-0.011852 + 0.093632i$ \\
        $0.9999$ & $2$ & $2$ & $0$ & $0.993234 - 0.007051i$ & $\phantom{-}1.501539 + 0.944334i$ \\
        $0.9999$ & $2$ & $2$ & $1$ & $0.993222 - 0.021193i$ & $\phantom{-}1.504794 + 0.946068i$ \\
        $0.9999$ & $2$ & $2$ & $2$ & $0.993112 - 0.049388i$ & $\phantom{-}1.513023 + 0.954197i$
        \\
        \hline
    \end{tabular*}
    \label{tab:QNM-comp-2}
\end{table*}
}

We find that our monodromy eigenvalues are consistent with those computed in Ref.~\cite{CastETC13b}. Most values agree with a fractional difference $\lesssim 10^{-5}$, which is approximately the level of precision to which the data are reported in Ref.~\cite{CastETC13b}. However, our results significantly differ at the level of $\sim 10^{-2}$ in two instances: $(a/M, l,m,n) = (0,0,0,2)$ and $(a/M, l,m,n) = (0.2,0,0,0)$. In the first case, the disagreement is reduced to $\sim 10^{-5}$ if we replace the dimensionless frequency $M\omega^\mathrm{Schw}_\mathrm{QNM}=0.075742-0.600080i$ given in Table B1 of Ref.~\cite{CastETC13b} with the slightly more accurate quasinormal mode frequency $0.075742-0.601080i$. Therefore, we use both frequencies in Table \ref{tab:QNM-comp}. For $(a/M, l,m,n) = (0.2,0,0,0)$, the source of the disagreement is less clear. While our comparison does indicate which result is more accurate, we find that our value $\nu^{(1)}_\infty$ is much closer to the monodromy eigenvalues for neighboring spin values, i.e., $(a/M, l,m,n) = (0,0,0,0)$ and $(a/M, l,m,n) = (0.4,0,0,0)$. Therefore, our value is consistent with nearby and verified results. Furthermore, we find that, in the near-extremal limit $a/M \gtrsim 0.9999$, some of the frequencies computed by \texttt{qnm} differ from those used in Ref.~\cite{CastETC13b}, and thus lead to slightly different monodromy eigenvalues, as evidenced in Table \ref{tab:QNM-comp-2}.

\begin{figure}
    \centering
    \includegraphics[width=0.48\linewidth]{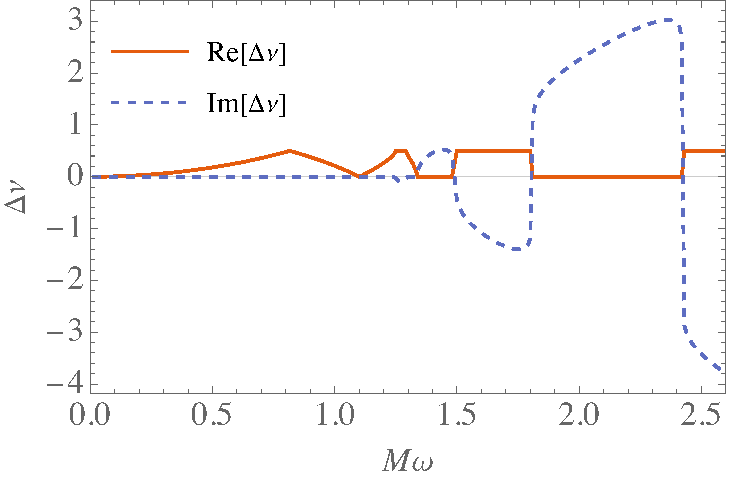}
    \hfill
    \includegraphics[width=0.5\linewidth]{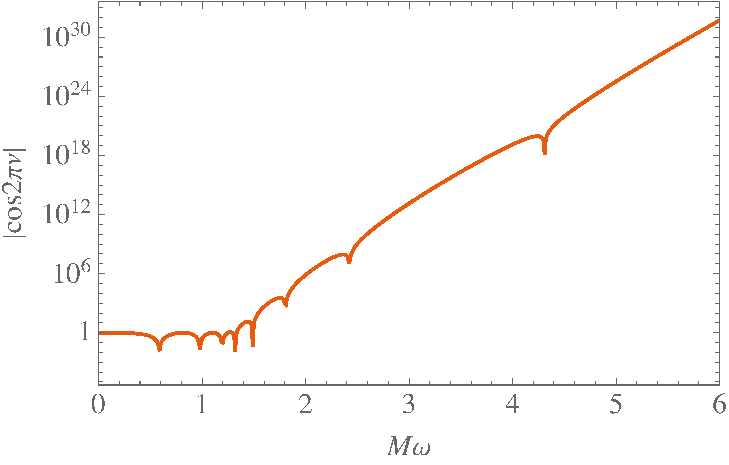}
    \caption{The monodromy eigenvalue as a function of (normalized) frequency $M\omega$ for the Teukolsky parameters $(s,l,m,\chi)=(-2,5,2,0.9)$. The left plot tracks the frequency evolution of $\nu$, while the right plot tracks $\cos 2\pi\nu$. While $\cos 2\pi\nu$ remains real for all real frequencies, $\nu$ jumps on and off the real-axis as we increase $M\omega$.}
    \label{fig:monodromy_sm2_l5_m2}
\end{figure}

\begin{figure}
    \centering
    \includegraphics[width=0.48\linewidth]{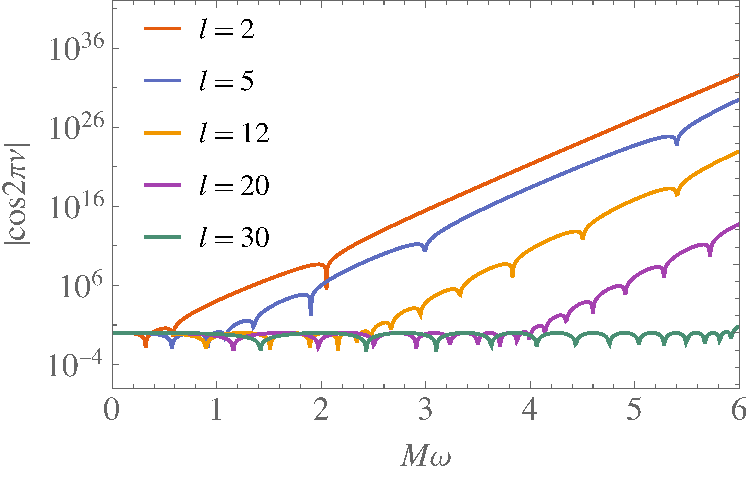}
    \includegraphics[width=0.48\linewidth]{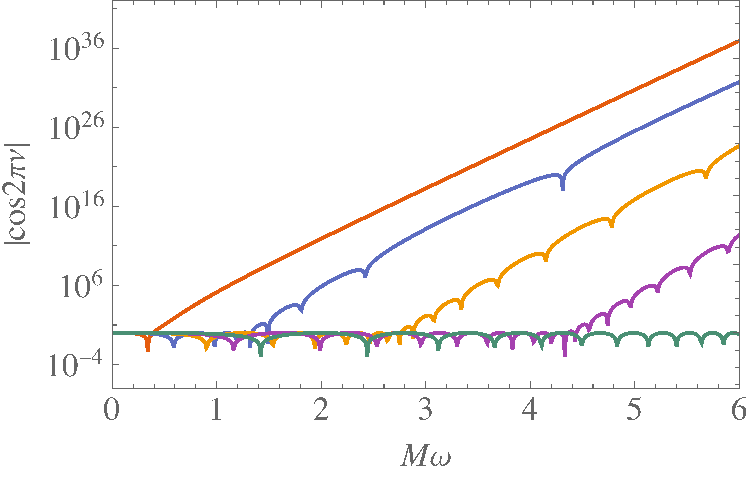}
    \caption{The monodromy eigenvalue as a function of (normalized) frequency $M\omega$ for the fixed Teukolsky parameters $(s,m)=(-2,2)$. The left plot demonstrates the effect of varying $l$ with $\chi=0.1$ fixed, while $\chi=0.9$ on the right.}
    \label{fig:monodromy_sm2_l_variation}
\end{figure}

\begin{figure}
    \centering
    \includegraphics[width=0.48\linewidth]{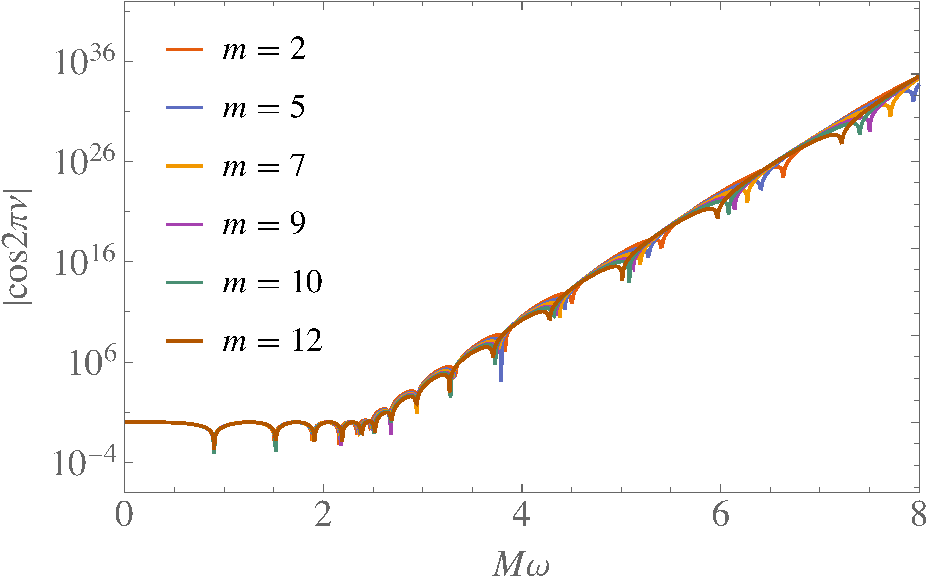}
    \includegraphics[width=0.48\linewidth]{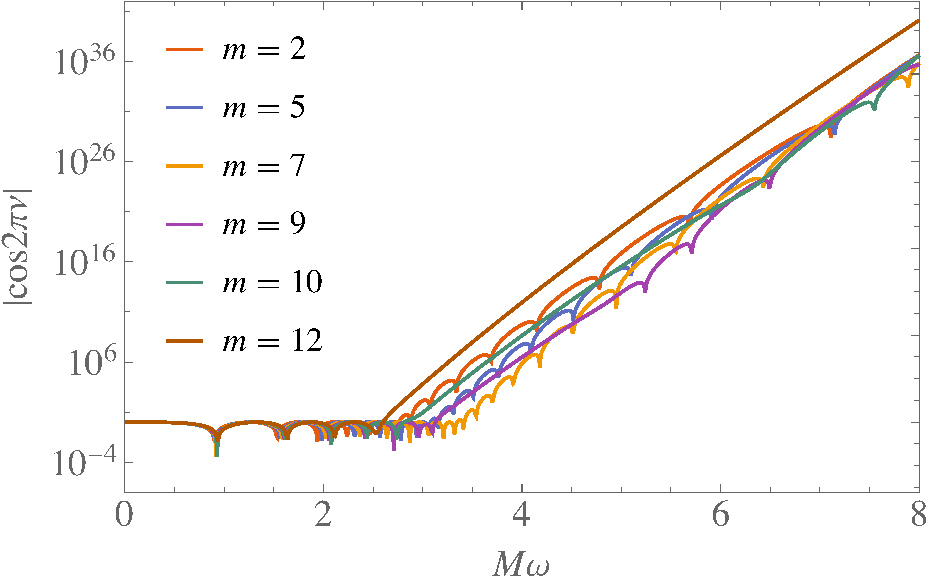}
    \caption{The monodromy eigenvalue as a function of (normalized) frequency $M\omega$ for the fixed Teukolsky parameters $(s,l)=(-2,12)$. The left plot demonstrates the effect of varying $m$ with $\chi=0.1$ fixed, while $\chi=0.9$ on the right.}
    \label{fig:monodromy_sm2_m_variation}
\end{figure}

Next, in Figure \ref{fig:monodromy_sm2_l5_m2}, we plot $\Delta\nu$ (left) and $\cos 2\pi\nu$ (right) as functions of (dimensionless) frequency $M\omega$ for the Teukolsky parameters $(s, l, m, \chi) = (-2, 5, 2, 0.9)$. As recognized in previous works (e.g., \cite{FujiTago05}), $\nu$ evolves on and off the real axis as the frequency increases. This is one reason why root-finding methods have struggled to efficiently compute $\nu$: it is not always clear where in the complex plane the zeros of Eq.~\eqref{eqn:CFnu2} are located for arbitrary values of $(s, l, m, \chi,\omega)$. While $\nu$ can be complex, we observe in our numerical calculations that $\cos2\pi\nu$ is always real for the radial Teukolsky equation when $(s, l, m, \chi, \omega)$ are real.

The evolution of $\cos2\pi\nu$ is also highly dependent on the value of $l$. In Figure \ref{fig:monodromy_sm2_l_variation}, we plot $\cos 2\pi\nu$ as a function of $M\omega$ for the Teukolsky parameters $(s, m, \chi) = (-2, 2, 0.1)$ (left) and $(s, m, \chi) = (-2, 2, 0.9)$ (right) but with varying values of $l$. As we increase $l$ or $\chi$, $\nu$ remains on the real axis over a larger range of frequencies. On the other hand, in Figure \ref{fig:monodromy_sm2_m_variation} we plot $\cos 2\pi\nu$ versus $M\omega$ for the Teukolsky parameters $(s, l, \chi) = (-2, 12, 0.1)$ (left) and $(s, l, \chi) = (-2, 12, 0.9)$ (right) but with varying values of $m$. Because $m$ only appears in the Teukolsky equation through the combination $m\chi$, the $m$-dependence is very weak at low spin values (see the left panel of Fig.~\ref{fig:monodromy_sm2_m_variation}), while at higher spins varying $m$ can either suppress or enhance the critical frequency at which $\cos2\pi\nu$ exponentially grows with $M\omega$ (see the right panel of Fig.~\ref{fig:monodromy_sm2_m_variation}). However, the effect is not as dramatic as increasing $l$, and the dependence of $\cos2\pi\nu$ on $m\chi$ is much more complicated. For example, as we initially increase $m$, $\nu$ remains real (e.g., $|\cos2\pi\nu| \leq 1$) for a larger range of frequencies in Fig.~\ref{fig:monodromy_sm2_m_variation}. Then, the trend reverses for $m \geq 9$, and $\cos2\pi\nu$ exponentially grows at lower and lower frequencies. This suggests that the value of $\cos2\pi\nu$ is primarily impacted by the values of $l$ (more specifically the spheroidal eigenvalue $\lambda^T$) and $\omega$, while the values of $\chi$ and $m\chi$ have subdominant effects.

\subsection{Numerical stability of the monodromy approach}
\label{sec:stability}

\begin{figure}
    \centering
    \includegraphics[height=2.05in]{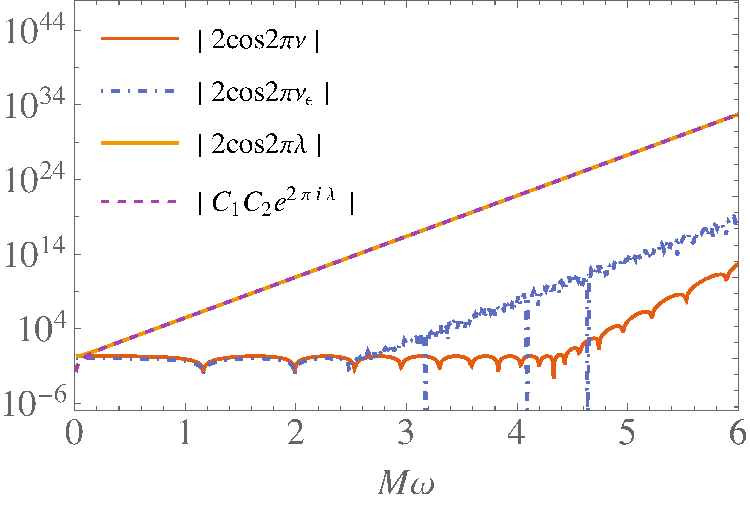}
    \hfill
    \includegraphics[height=2.05in]{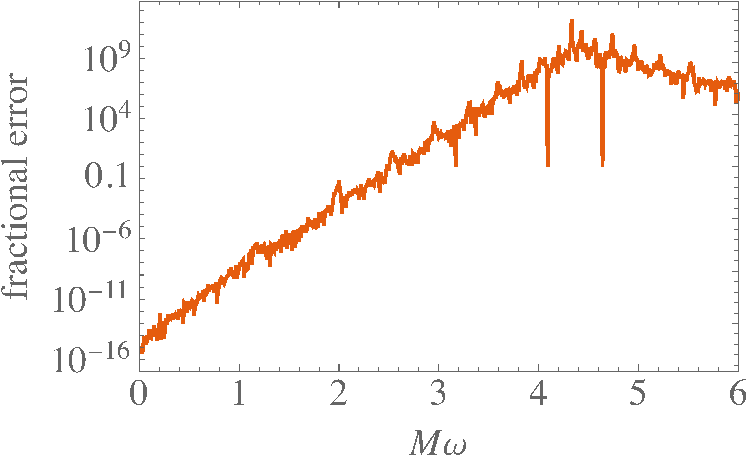}
    \caption{The left panel plots $2\cos 2\pi \nu$ (red solid line), $2\cos2\pi\lambda = 2\cos2\pi\xi$ (yellow solid line), and $C_1 C_2 e^{2\pi i \lambda}$ (dashed purple line) as functions of (normalized) frequency $M\omega$ for the fixed Teukolsky parameters $(s,l,m,\chi)=(-2,20,2,0.9)$. A machine-precision calculation of the monodromy eigenvalue $2\cos 2\pi \nu_\epsilon$ (dot-dashed blue line) is also plotted to demonstrate the effects of catastrophic cancellation. The right panel then displays the fractional error between $\cos 2\pi \nu$ and $\cos 2\pi \nu_\epsilon$.}
    \label{fig:nuCancellation}
\end{figure}

\begin{figure}
    \centering
    \includegraphics[height=2.05in]{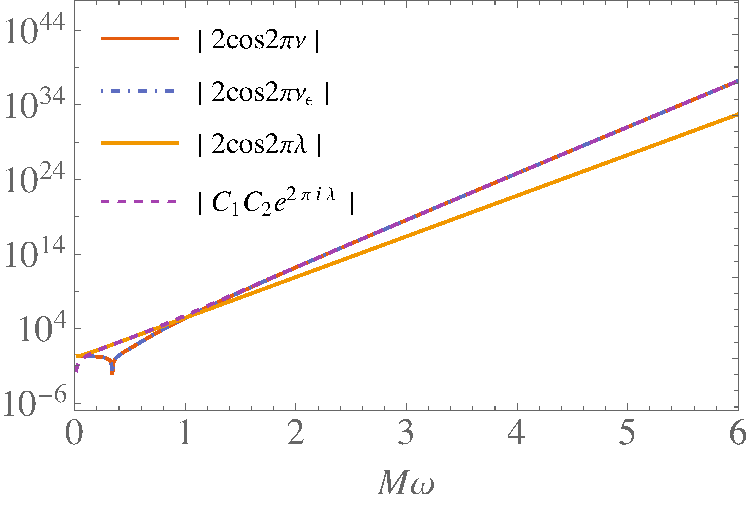}
    \hfill
    \includegraphics[height=2.05in]{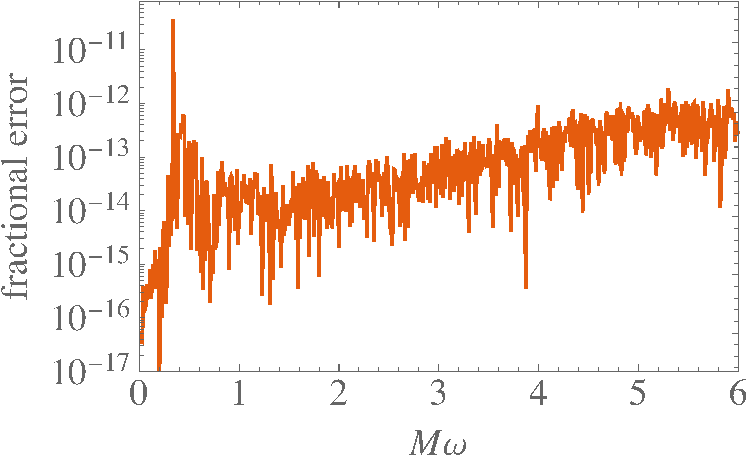}
    \caption{The same as Figure \ref{fig:nuCancellation} but for the Teukolsky parameters $(s,l,m,\chi)=(-2,2,2,0.9)$.}
    \label{fig:nuCancellationL2}
\end{figure}

One numerical limitation of this monodromy approach is that Eq.~\eqref{eqn:monodromyEigenvalues} suffers from catastrophic cancellation for larger values of both $l$ and $M\omega$. In the left panel of Figure \ref{fig:nuCancellation}, we plot $2\cos2\pi\lambda$ (yellow solid line), $C_1 C_2 e^{2\pi i \lambda}$ (dashed purple line), and $2\cos 2\pi \nu$ (red solid line) as functions of $M\omega$ for the Teukolsky parameters $(s,l,m,\chi)=(-2, 20, 2, 0.9)$. As the frequency increases, $2\cos2\pi\lambda$ and $C_1 C_2 e^{2\pi i \lambda}$ grow exponentially, while the monodromy eigenvalue remains bounded by $|\cos 2\pi\nu| \leq 1$ up until $M\omega \sim 4.5$. Therefore $\cos 2\pi\nu$ can only be extracted after subtracting off several orders of magnitude between $2\cos2\pi\lambda$ and $C_1 C_2 e^{2\pi i \lambda}$. We can estimate the precision loss by the fractional difference $|1-\cos2\pi\lambda/\cos 2\pi \nu| \approx e^{4\pi\omega}$ for $\cos 2\pi \nu \sim 1$. For frequencies $M\omega \gtrsim 2.75$, one loses over 15 digits of precision due to catastrophic cancellations in \eqref{eqn:monodromyEigenvalues}.

To highlight the impact of this catastrophic cancellation, we plot two different numerical values for the monodromy eigenvalue in Fig.~\ref{fig:nuCancellation}. The first value, which we denote as $\cos2\pi\nu$ (solid red line), is accurately calculated using arbitrary-precision arithmetic. The second value, which we refer to as $\cos2\pi\nu_\epsilon$ (dot-dashed blue line), is calculated using machine-precision arithmetic, leading to inaccurate results at higher frequencies. The fractional error between these two calculations, i.e., $|1-\cos2\pi\nu/\cos2\pi\nu_\epsilon|$, is plotted in the right panel of Fig.~\ref{fig:nuCancellation}. As expected, the fractional error becomes larger than unity for frequencies $M\omega \gtrsim 2.75$, indicating that $\cos2\pi\nu_\epsilon$ is completely dominated by numerical noise. This is also evident in the left panel of Fig.~\ref{fig:nuCancellation}: $\cos2\pi\nu_\epsilon$ grows exponentially with the numerical noise for the same range of frequencies.

The degree of catastrophic cancellation is also heavily impacted by the value of $l$, as one might expect based on Fig.~\ref{fig:monodromy_sm2_l_variation} and the discussion in Sec.~\ref{sec:results}. In Figure \ref{fig:nuCancellationL2} we repeat this analysis for $l=2$. We observe a much smaller degree of cancellation, because, for smaller values of $l$, $|\cos 2\pi\nu|$ is much closer in magnitude to $\cos 2\pi\lambda$ across frequency space. In other words, little cancellation occurs in \eqref{eqn:monodromyEigenvalues}. On the other hand, as demonstrated in Figs.~\ref{fig:monodromy_sm2_l_variation} and \ref{fig:nuCancellationL2}, $\cos 2\pi\nu$ remains bounded over a larger range of frequencies for higher $l$-modes. Thus the cancellations grow to be more and more catastrophic as both $M\omega$ and $l$ increase.

To partially circumvent this issue at large $l$ values, we make use of the asymptotic behavior of $\cos 2\pi \nu$. In particular, when $|\cos 2\pi \nu| \leq 1$ but $\lambda^T \gg 1$, we expect that $\cos 2\pi \nu \sim - \cos 2\pi \sqrt{\lambda^T}$. Defining, $\lambda_C = \lambda^T + s(s+1)$, we form the ansatz,
\begin{subequations} \label{eqn:monodromyFit}
    \begin{align} 
        \cos 2\pi \nu &\sim -\cos 2\pi\left[\lambda_C^{1/2} + \nu_{1} \lambda_C^{-1/2} + \nu_{3} \lambda_C^{-3/2} + \nu_{5} \lambda_C^{-5/2} + \nu_{7} \lambda_C^{-7/2} + O\left( \lambda_C^{-9/2}\right)\right].
    \end{align} 
\end{subequations}
We then numerically calculate $\cos 2\pi \nu$ at large values of $\lambda^T$ and extract the following coefficients,
\begin{subequations} \label{eqn:monodromyFitParameters}
    \begin{align} 
        \nu_{1} &= \frac{1}{8} + \left(m \chi \right)\frac{\epsilon}{2} - \frac{1}{4}\left({15} +{\chi^2}\right)\frac{\epsilon^2}{4},
        \\
        \nu_{3} &= -\frac{1}{128} - m\chi\left(\frac{1}{8}- s^2 \right) \frac{\epsilon}{2}+  \frac{1}{2}\left[\frac{13}{16} - 3 s^2 + \left(\frac{3}{16} - \frac{3m^2}{2} - s^2\right)\chi^2\right]\frac{\epsilon^2}{4} 
        \\ \notag
        & \qquad \qquad \qquad \qquad \qquad \qquad \qquad
        + \frac{m \chi}{4}\left(35 + \chi^2\right) \frac{\epsilon^3}{8}- \frac{1}{32}\left(\frac{1155}{2} - {35}\chi^2 + \frac{3 \chi^4}{2}\right)\frac{\epsilon^4}{16},
    \end{align} 
\end{subequations}
while $\nu_5$ and $\nu_7$ are given in Appendix \ref{app:expansion}.

\begin{figure}
    \centering
    \includegraphics[width=0.98\linewidth]{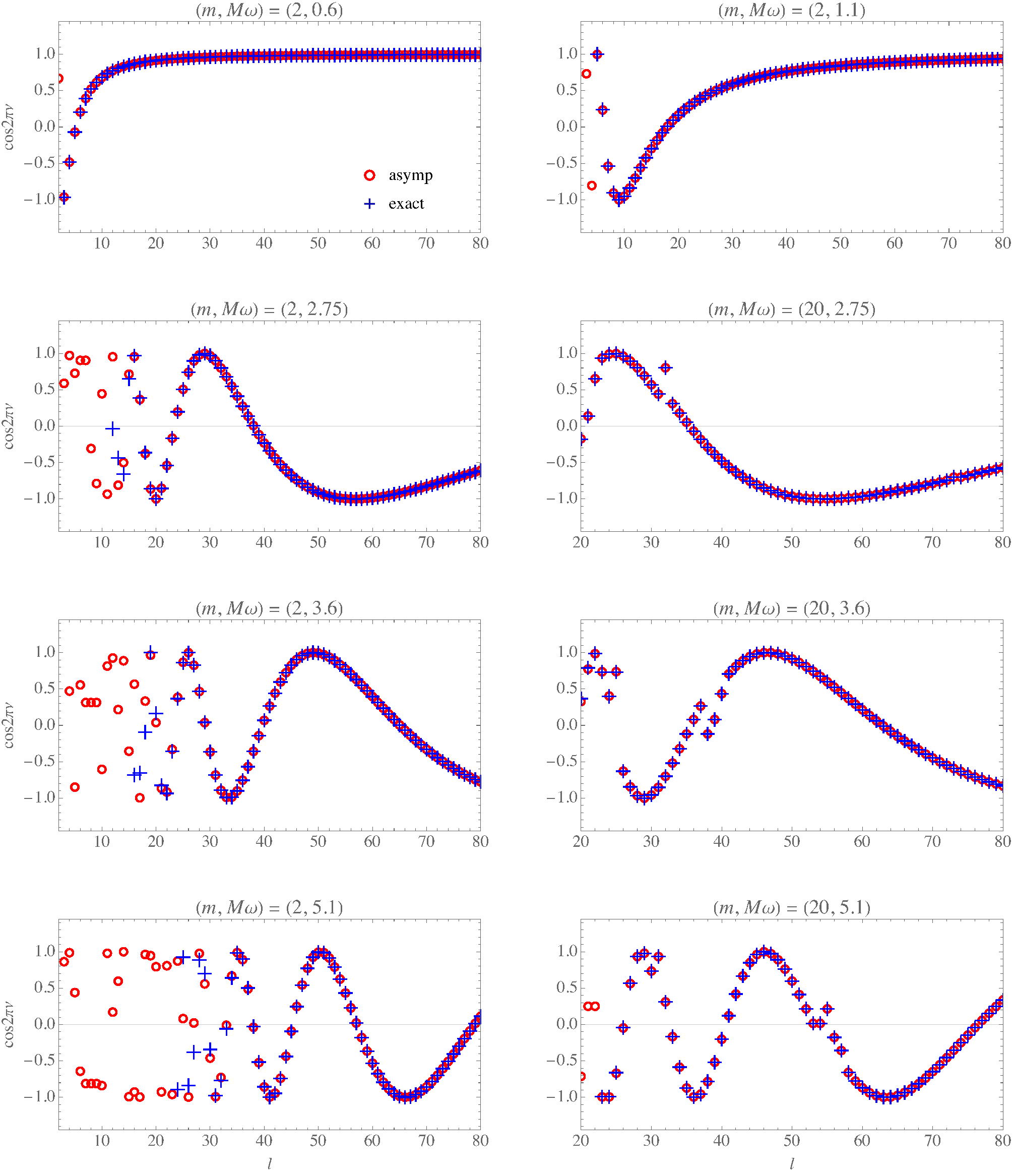}
    \caption{Comparing exact calculations of $\cos2\pi\nu$ (blue crosses) to the values predicted by the asymptotic expansion in \eqref{eqn:monodromyFit} (red circles) as functions of $l \geq |m|$ for various values of $(m,M\omega)$ but with $(s,\chi)=(-2, 0.9)$ fixed.}
    \label{fig:asympComparison}
\end{figure}

\begin{figure}
    \centering
    \includegraphics[width=0.98\linewidth]{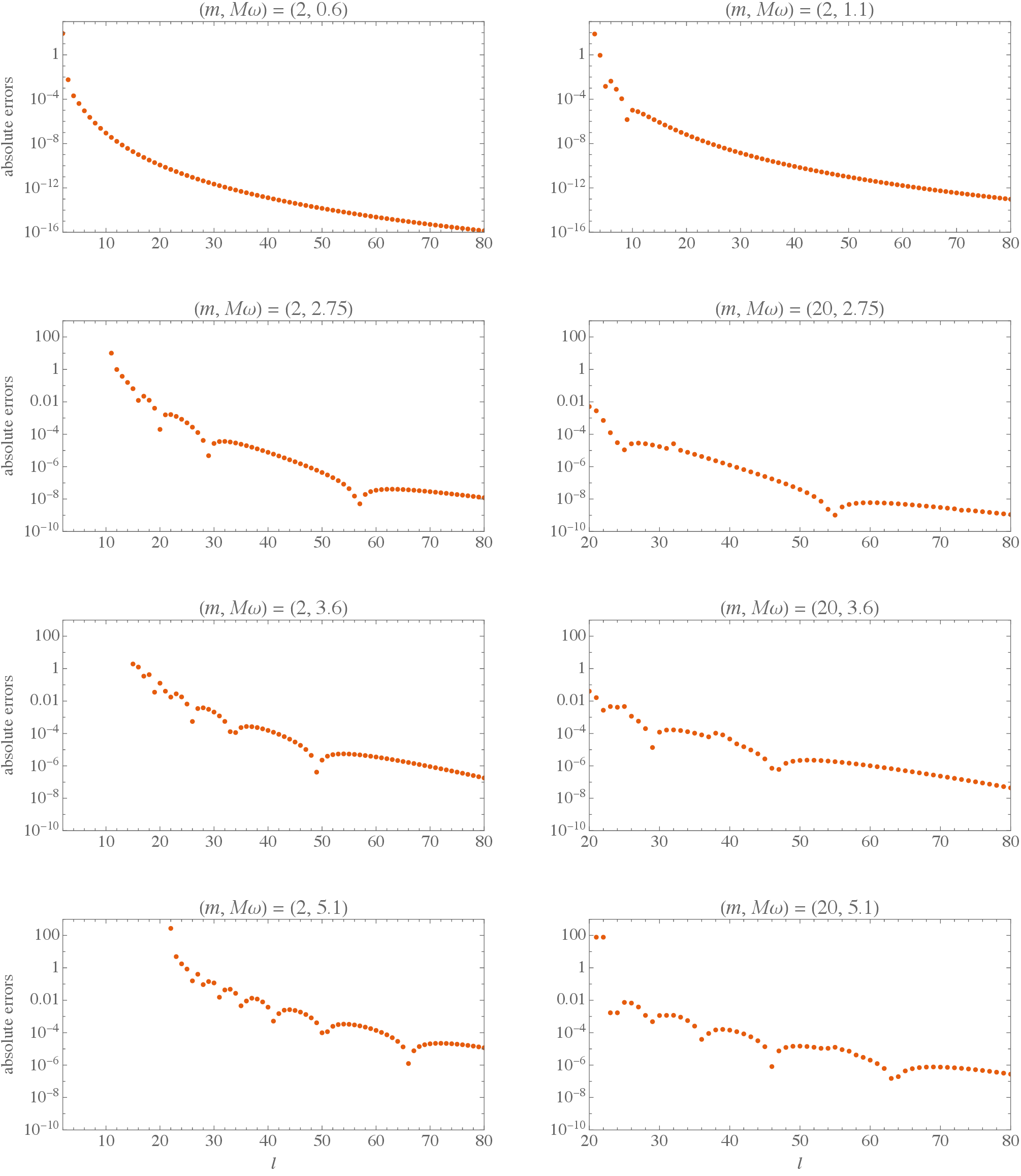}
    \caption{The absolute error between an exact calculation of $\cos2\pi\nu$ and the asymptotic expansion in \eqref{eqn:monodromyFit} as a function of $l$ for the same values of $(s,l,m,\chi,M\omega)$ as displayed in Fig.~\ref{fig:asympComparison}.}
    \label{fig:asympErrors}
\end{figure}

There are several limitations to this expansion. First of all, the series is asymptotic and not guaranteed to converge for arbitrary values of $\lambda_C$ and $\epsilon = 2M\omega$. In particular, the terms have the frequency scaling $\nu_k \sim \epsilon^{k+1}$. Consequently, $\nu_{2k+1}$ will not decay as $k\rightarrow \infty$ for large enough values of $\epsilon$. Additionally, the expansion assumes $|\cos2\pi\nu| \leq 1$, but we do not know \emph{a priori} whether or not this is true for arbitrary values of $(s,l,m,\chi,\omega)$. However, when $\nu_{2k+3} \gtrsim \nu_{2k+1} \lambda_C$, this indicates the asymptotic expansion is breaking down and that $\cos2\pi\nu$ is growing exponentially with frequency rather than oscillating with the value of $\lambda_C$. In Figure \ref{fig:asympComparison}, we compare the asymptotic expansion in \eqref{eqn:monodromyFit} (red circles) to an ``exact" calculation of $\cos2\pi\nu$ (blue crosses) via Eq.~\eqref{eqn:monodromyEigenvalues}. We plot both the exact and asymptotic values as functions of $l\geq|m|$ for various combinations of $m$ and $M\omega$. Taking the ``exact" calculation to be the true value, in Figure \ref{fig:asympErrors}, we plot the absolute errors of the asymptotic results. As we expect, the asymptotic expansion is most accurate at small frequencies and large values of $l$. At the frequency $M\omega \sim 2.75$---where we expect to lose all machine-precision information due to catastrophic cancellation---the asymptotic expansion is able to recover $\cos2\pi\nu$ within a few digits of accuracy for $l \gtrsim 16$. Thus the asymptotic expansion struggles at low $l$ but high $M\omega$, where $\cos2\pi\nu$ is transitioning to its exponential growth with frequency. Ultimately, this asymptotic approach may work better as an initial guess for the value of $\nu$, which can be combined with previous root-finding algorithms that rely on the MST constraint equations \eqref{eqn:CFnu} or \eqref{eqn:CFnu2} to extract $\nu$. 

Alternatively, one could resum or reexpand \eqref{eqn:monodromyFit} to also take into account the asymptotic behavior $\nu \sim -i\epsilon$ for $M\omega\gg 1$. One choice is to expand in $(\lambda_C - \epsilon^2)^{-1}$. This would lead to the correct behavior in the two asymptotic limits $\lambda_C \rightarrow \infty$ and $\epsilon \rightarrow \infty$, but the expansion would break down for $\lambda_C = \epsilon^2$. Thus, one would still require a different series representation for the transition between the two regimes. We leave further investigations of these expansions for future work.

\section{Conclusion}
\label{sec:conclusion}

In this work we demonstrated that MST's renormalized angular momentum parameter $\nu$ is not merely an auxilliary parameter, but is directly related to the monodromy eigenvalues of the irregular singular point of the radial Teukolsky equation in Kerr spacetime. To establish this relationship, we first recognized that the Teukolsky solutions $R^\nu_C$ and $R^{-\nu-1}_C$ described in Eq.~\eqref{eqn:RnuC} [and likewise $R^\nu_0$ and $R^{-\nu-1}_0$ in Eq.~\eqref{eqn:Rnu0}] diagonalize the monodromy matrix at infinity and provide a natural basis for studying the behavior of the Teukolsky solutions near this singular point. In Sec.~\ref{sec:numerical} we outlined practical numerical methods for obtaining $\nu$ from the Stokes multipliers and monodromy eigenvalues of the Teukolsky equation by combining Eqs.~\eqref{eqn:monodromyEigenvalues}, \eqref{eqn:C1C2compact}, and \eqref{eqn:nuToNuInf}. Using these methods, we then calculated the renormalized angular momentum across the parameter space and found that $\cos2\pi\nu$ is always real when the Teukolsky parameters $(s,l,m,\chi,M\omega)$ are real. This is in contrast to $\nu$, which can be real or complex even when $(s,l,m,\chi,M\omega)$ are real. We also highlighted limitations to this monodromy approach, particularly issues of catastrophic cancellation when evaluating Eq.~\eqref{eqn:monodromyEigenvalues}, and proposed potential methods for mitigating these problems, which make use of new asymptotic expansions of $\cos2\pi\nu$ in \eqref{eqn:monodromyFit}.

Naturally, one can use these numerical methods to calculate $\nu$ and evaluate the MST series solutions. Alternatively, due to the relationship between the Teukolsky and confluent Heun equations, one can also construct radial Teukolsky solutions by leveraging software packages that now include confluent Heun solutions within their special function libraries, such as \textsc{Mathematica}'s \texttt{HeunC} function. Combining these special functions with the MST amplitudes in Eqs.~\eqref{eqn:MSTamplitudes} and \eqref{eqn:MSTamplitudes2}---all of which depend on $\nu$---one can obtain any independent set of radial Teukolsky functions. 

One can also make use of the MST amplitudes (see App.~\ref{app:MST}) and $\nu$ to construct scattering data in Kerr spacetime, such as greybody factors or tidal Love numbers. Furthermore, over the past decade, there has been a flurry of research connecting monodromy data, conformal blocks, supersymmetric gauge theory, and the Painlevé
VI transcendent to obtain analytic expansions of black hole quasinormal modes and scattering amplitudes (e.g., \cite{CastETC13a,CarnNova15,NovaETC19,BautETC24,AminArna24}). Connecting $\nu$ to monodromy also connects it to these various approaches. For example, our results verify that Eq.~(3.6) in Ref.~\cite{BautETC24} (i.e., $a = -1/2-\nu$) is exact.\footnote{Ref.~\cite{BautETC24} only establishes this equality to 9th post-Minkowskian order.} Thus this work further elucidates the rich relationship between the MST solutions and scattering theory.

\begin{acknowledgments}
The material is based upon work supported by NASA under award number 80GSFC21M0002. This work was also supported by NSF Grant No.~PHY-1806447 to the University of North Carolina--Chapel Hill. The author also thanks A.C.~Ottiwell, B.~Wardell, M.~Casals, and C.R.~Evans for useful discussions. This work makes use of the Black Hole Perturbation Toolkit.
\end{acknowledgments}

\appendix

\section{Recurrence relations for expansions of confluent Heun functions}
\label{app:CHEdefinitions}

Solutions to the confluent Heun equation \eqref{eqn:che} can be approximated by the series expansion around $z=1$, as expressed in Eq.~\eqref{eqn:cheRegular}. The coefficients $\hat{a}_{j,k}$ in \eqref{eqn:cheRegular} satisfy the three-term recurrence relation \eqref{eqn:cheRegRecurrence} with,
\begin{subequations}
    \begin{align}
        A^\mathcal{H}_{j,k} &= \alpha_\mathrm{CH} +\varepsilon_\mathrm{CH} (n + \lambda^\mathcal{H}_j - 1),
        \\
        B^\mathcal{H}_{j,k} &= n^2 + n(\gamma_\mathrm{CH} + \delta_\mathrm{CH} + \epsilon_\mathrm{CH} + 2 \lambda^\mathcal{H}_j - 1)
        \\ \notag
        & \qquad \qquad \qquad + \lambda^\mathcal{H}_j(\gamma_\mathrm{CH} + \delta_\mathrm{CH} + \varepsilon_\mathrm{CH} + \lambda^\mathcal{H}_j - 1) - q_\mathrm{CH} + \alpha_\mathrm{CH},
        \\
        C^\mathcal{H}_{j,k} &= (n + 1 + \lambda^\mathcal{H}_j)(n + \delta_\mathrm{CH}  + \lambda^\mathcal{H}_j),
    \end{align}  
\end{subequations}
Similarly confluent Heun solutions are asymptotic to series expansions around $z=\infty$, as given in Eq.~\eqref{eqn:cheIrregular}. The coefficients $\hat{b}_{j,k}$ in \eqref{eqn:cheIrregular} satisfy the three-term recurrence relation \eqref{eqn:cheRecurrence} with,
\begin{subequations}
    \begin{align}
        A^\mathcal{I}_{j,k} &= -[\alpha_\mathrm{CH} +\mu^\mathcal{I}_j  (\gamma_\mathrm{CH} +\delta_\mathrm{CH} +2 k-2)+(k-1) \varepsilon_\mathrm{CH} ] 
        \\ \notag
        & \qquad \times [\alpha_\mathrm{CH} -\gamma_\mathrm{CH}  (\mu^\mathcal{I}_j
       +\varepsilon_\mathrm{CH} )+\delta_\mathrm{CH}  \mu^\mathcal{I}_j +k (2 \mu^\mathcal{I}_j +\varepsilon_\mathrm{CH} )],
        \\ \notag
        B^\mathcal{I}_{j,k} &= \mu^\mathcal{I}_j  \varepsilon_\mathrm{CH}  \left(\gamma_\mathrm{CH}-(\gamma_\mathrm{CH}
       +\delta_\mathrm{CH} )^2 +\delta_\mathrm{CH} (1+ \varepsilon_\mathrm{CH})-2 k (\gamma_\mathrm{CH} +\delta_\mathrm{CH} -3 \varepsilon_\mathrm{CH}
       -2k-2)-4 q_\mathrm{CH}\right)
        \\ 
        &\qquad -(\mu^\mathcal{I}_j)^2 \big((\gamma_\mathrm{CH} +\delta_\mathrm{CH} -2) (\gamma_\mathrm{CH} +\delta_\mathrm{CH} )-4 \big(\delta_\mathrm{CH}
        \varepsilon_\mathrm{CH} +k^2+3 k \varepsilon_\mathrm{CH} +k\big)+4 q_\mathrm{CH}\big)
       \\  \notag
       &\qquad \qquad +\alpha_\mathrm{CH}  \left(\varepsilon_\mathrm{CH}  (-\gamma_\mathrm{CH} -\delta_\mathrm{CH} +2 k+4 \mu^\mathcal{I}_j +1)+4 \mu^\mathcal{I}_j  (k+\mu^\mathcal{I}_j
       )+2 \mu^\mathcal{I}_j +\varepsilon_\mathrm{CH} ^2\right)
       \\  \notag
       & \qquad \qquad \qquad +4 (\mu^\mathcal{I}_j)^3 (\delta_\mathrm{CH} +2 k)+\varepsilon_\mathrm{CH} ^2 (k (-\gamma_\mathrm{CH} -\delta_\mathrm{CH}
       +k+\varepsilon_\mathrm{CH} +1)-q_\mathrm{CH}) +\alpha_\mathrm{CH}^2,
        \\
        C^\mathcal{I}_{j,k} &= -(k+1) (2 \mu^\mathcal{I}_j +\varepsilon_\mathrm{CH} )^3.
    \end{align}
\end{subequations}

\section{MST methods}
\label{app:MST}

The recurrence relation \eqref{eqn:coeff} for the MST series coefficients $f^\nu_n$ is defined in terms of recurrence coefficients $\alpha^\nu_n = A_{\nu+n}$, $\beta^\nu_n = B_{\nu+n}$, and $\gamma^\nu_n = A_{-\nu-n-1}$, where,
\begin{subequations}
    \begin{align}
        A_L &= \frac{i\epsilon\kappa(L+1+\bar{\xi})(L+1+\xi)(L+1+i\tau)}{(L+1)(2L+3)},
        \\
        B_L &= L(L+1) -\lambda_C + \epsilon^2 + \epsilon\kappa\tau \left[1+ \frac{\xi \bar{\xi}}{L(L+1)}\right].
    \end{align}  
\end{subequations}
Recall that $\lambda_C = \lambda^T + s(s+1)$. 

We also define the following coefficients to condense notation when relating the different homogeneous solutions in Eqs.~\eqref{eqn:RinToRnu0} and \eqref{eqn:MSTamplitudes},
\begin{subequations}
    \begin{align}
        B^\nu_0 &= \frac{\Gamma(1+\bar{\xi}+i\tau)\Gamma(1+\nu-\bar{\xi})\Gamma(1+\nu-i\tau)}{\Gamma(1-\bar{\xi}-i\tau)\Gamma(1+\nu+\bar{\xi})\Gamma(1+\nu+i\tau)},
        \\ \label{eqn:Knu}
        K^\nu &= e^{i\epsilon \kappa} (\epsilon\kappa)^{s-\nu} 2^{-\nu} \left(\sum_{n=r}^\infty C_{n,n-r}\right)\left(\sum_{n=-\infty}^r D_{n,r-n}\right)^{-1},
        \\
        A^\nu_+ &= 2^{-1+\xi} e^{\frac{\pi i}{2}(\nu+1-\xi)} \frac{\Gamma(\nu+1-\xi)}{\Gamma(\nu+1+\xi)} \sum_{n=-\infty}^\infty f^\nu_n,
        \\
        A^\nu_- &= 2^{-1-\xi} e^{-\frac{\pi i}{2}(\nu+1+\xi)}  \sum_{n=-\infty}^\infty (-1)^n \frac{(\nu+1+\xi)_n}{(\nu+1-\xi)_n} f^\nu_n,
    \end{align}
\end{subequations}
where,
\begin{align}
    C_{n,j} &= (\epsilon\kappa)^{-n+j}\frac{\Gamma(1-\bar{\xi}-i\tau)\Gamma(2n+2\nu+1)}{\Gamma(n+\nu+1-\bar{\xi})\Gamma(n+\nu+1-i\tau)}\frac{(-n-\nu-\bar{\xi})_j(-n-\nu-i\tau)_j}{(-2n-2\nu)_j}\frac{f^\nu_n}{j!},
    \\
    D_{n,j} &= (-1)^n (2i)^{n+j} \frac{\Gamma(n+\nu+1-\xi)}{\Gamma(2n+2\nu+1)}\frac{(\nu+1+\xi)_n}{(\nu+1-\xi)_n} \frac{(n+\nu+1-\xi)_j}{(2n+2\nu+2)_j}\frac{f^\nu_n}{j!},
\end{align}
and $K^\nu$ can be computed using any integer $r$ in Eq.~\eqref{eqn:Knu}.

By matching to the asymptotic behavior of the MST solutions to Eq.~\eqref{eqn:teukAsymp2}, we also provide their reflection and incidence scattering amplitudes,
\begin{subequations} \label{eqn:MSTamplitudes2}
    \begin{align}
        \mathcal{R}^\mathrm{in,inc} &= \left[K^\nu - i e^{-i\pi\nu} \frac{\sin\pi(\nu - \xi)}{\sin\pi(\nu + \xi)} K^{-\nu-1} \right] \mathcal{R}^\mathrm{down,trans},
        \\
        \mathcal{R}^\mathrm{in,ref} &= \left[K^\nu + i e^{i\pi\nu} K^{-\nu-1} \right] \mathcal{R}^\mathrm{up,trans},
        \\
        \mathcal{R}^\mathrm{out,inc} &= \left[B^\nu K^\nu - i e^{-i\pi\nu} \frac{\sin\pi(\nu - \xi)}{\sin\pi(\nu + \xi)} B^{-\nu-1} K^{-\nu-1} \right] \mathcal{R}^\mathrm{down,trans},
        \\
        \mathcal{R}^\mathrm{out,ref} &= \left[B^\nu K^\nu + i e^{i\pi\nu}B^{-\nu-1} K^{-\nu-1} \right] \mathcal{R}^\mathrm{up,trans},
        \\
        \mathcal{R}^\mathrm{up,inc} &= \frac{D^{-\nu-1}}{B^\nu \sin2\pi\nu}\left[\frac{\sin\pi(\nu-\xi) e^{-i\pi(\nu+\xi)}}{K^\nu} + \frac{\sin\pi(\nu+\xi) ie^{-i\pi\xi}}{K^{-\nu-1}} \right]  \mathcal{R}^\mathrm{out,trans},
        \\
        \mathcal{R}^\mathrm{up,ref} &= \frac{1}{\sin2\pi\nu}\left[\frac{{D^\nu}\sin\pi(\nu-\xi) e^{-i\pi(\nu+\xi)}}{K^\nu} - \frac{D^{-\nu-1}\sin\pi(\nu+\xi) ie^{-i\pi\xi}}{K^{-\nu-1}} \right]  \mathcal{R}^\mathrm{in,trans},
        \\
        \mathcal{R}^\mathrm{down,inc} &= \frac{D^{-\nu-1}\sin\pi(\nu+\xi)}{B^\nu \sin2\pi\nu}\left[\frac{e^{i\pi(\nu-\xi)}}{K^\nu} - \frac{ ie^{-i\pi\xi}}{K^{-\nu-1}} \right]  \mathcal{R}^\mathrm{out,trans},
        \\
        \mathcal{R}^\mathrm{down,ref} &= \frac{\sin\pi(\nu+\xi)}{\sin2\pi\nu}\left[\frac{{D^\nu}e^{i\pi(\nu-\xi)}}{K^\nu} + \frac{D^{-\nu-1} ie^{-i\pi\xi}}{K^{-\nu-1}} \right]  \mathcal{R}^\mathrm{in,trans},
    \end{align}
\end{subequations}
with,
\begin{align}
    D^\nu = \frac{B^{-\nu-1}}{B^{-\nu-1} - B^\nu} = -\frac{\sin\pi(\nu-\bar{\xi}) \sin\pi(\nu-i\tau)}{\sin2\pi\nu \sin\pi(\bar{\xi}+i\tau)}.
\end{align}

\section{Connection formulae}
\label{app:connection}

Consider a solution $\psi^\infty_{k}(z)$ to Eq.~\eqref{eqn:ode}.
Recall from Eq.~\eqref{eqn:irregularSeriesAsymp} that $\psi^\infty_{j}(z)$ is asymptotic to the series expansion $\hat{\psi}_j^\infty(z)$ [see Eq.~\eqref{eqn:irregularSeries}] in the wedge $\hat{S}_j$ of the complex domain [see Eq.~\eqref{eqn:Sj})]. Solutions in neighboring sectors $\{\hat{S}_{j+2},\hat{S}_{j+1},\hat{S}_j\}$ are then related via the connection formula,
\begin{align}
    \psi^\infty_{j+2}(z) = C_j \psi^\infty_{j+1}(z) + \psi^\infty_j(z),
\end{align}
where,
\begin{align}
    \psi^\infty_{j+2k}(z) = e^{2\pi i k \lambda^\infty_j} \psi^\infty_{j} (e^{-2\pi i k} z),
\end{align}
with $\lambda_j^\infty = \lambda^\infty_1$ if $j$ is odd and $\lambda^\infty_j = \lambda^\infty_2$ if $j$ is even. From this, one can derive \eqref{eqn:connectionFormulae}.

\section{Equivalence of Stokes multipliers}

We briefly demonstrate the Stokes multipliers associated with the MST solutions $R^\mathrm{up/down}(z)$ defined in Eqs.~\eqref{eqn:Rup} and \eqref{eqn:Rdown} are equivalent to the transformed radial functions $w^\mathrm{up/down}(z)$, which satisfy the confluent Heun equation \eqref{eqn:che}. Using the transformation defined in Eq.~\eqref{eqn:cheTransform}, along with connection formulae \eqref{eqn:connectionFormulae}, leads to the relations,
\begin{subequations} \label{eqn:mixedConnections}
    \begin{align}
        e^{-2\pi\epsilon} R^\mathrm{up}(e^{-2\pi i}\hat{z}) &= R^\mathrm{up}(\hat{z}) + C_1 R^\mathrm{down}(\hat{z})
        \\
        &= \hat{z}^{i\epsilon_-}(\hat{z}-1)^{i\epsilon_+} \left[w^\mathrm{up}(\hat{z}) + C_1 w^\mathrm{down}(\hat{z})\right],
        \\
        e^{-2\pi\epsilon} R^\mathrm{down}(e^{2\pi i}\hat{z}) &= R^\mathrm{down}(\hat{z}) - C_2 R^\mathrm{up}(\hat{z})
        \\
        &= \hat{z}^{i\epsilon_-}(\hat{z}-1)^{i\epsilon_+} \left[w^\mathrm{down}(\hat{z}) - C_2 w^\mathrm{up}(\hat{z})\right],
    \end{align}
\end{subequations}
where $C_1$ and $C_2$ are the Stokes multipliers associated with $R^\mathrm{up/down}(z)$. Furthermore, as we approach infinity, Eq.~\eqref{eqn:cheTransform} also yields,
\begin{align} \label{eqn:RinfToWinfGamma}
    R^\mathrm{up/down}(e^{\mp 2\pi i}\hat{z}) &= e^{\pm 2\pi\epsilon}\hat{z}^{i\epsilon_-} (\hat{z}-1)^{i\epsilon_+} e^{i\epsilon\kappa \hat{z}} w^\mathrm{up/down}(e^{\mp 2\pi i}\hat{z}), 
    &
    (z &\rightarrow \infty).
\end{align}
Combining Eq.~\eqref{eqn:mixedConnections} with \eqref{eqn:RinfToWinfGamma} then leads to,
\begin{align} \label{eqn:connectionVerify}
    w^\mathrm{up}(e^{-2\pi i}\hat{z}) &= w^\mathrm{up}(\hat{z}) + C_1 w^\mathrm{down}(\hat{z}),
    \\
    e^{-4\pi\epsilon} w^\mathrm{down}(e^{2\pi i}\hat{z}) &= w^\mathrm{down}(\hat{z}) - C_2 w^\mathrm{up}(\hat{z}),
\end{align}
which holds for all $z$, since $w^\mathrm{up/down}(e^{-2\pi i}\hat{z})$ and $w^\mathrm{up/down}(\hat{z})$ are all independent homogeneous solutions of \eqref{eqn:che}. Because Eq.~\eqref{eqn:connectionVerify} is equivalent to the connection equations \eqref{eqn:connectionFormulae}, $C_1$ and $C_2$ must also be the Stokes multipliers for $w^\mathrm{up/down}(z)$.

\section{Higher-order coefficient for asymptotic fit of the monodromy eigenvalue}
\label{app:expansion}

The higher-order fitting coefficients $\nu_5$ and $\nu_7$ for the asymptotic expansion of $\cos2\pi\nu$ in Eq.~\eqref{eqn:monodromyFit} are given by,
\begin{align}
    \nu_5 &= \sum_{n=0}^6 \nu_5^{(n)} \left(\,\frac{\epsilon}{2}\,\right)^n,
    &
    \nu_7 &= \sum_{n=0}^8 \nu_7^{(n)} \left(\,\frac{\epsilon}{2}\,\right)^n,
\end{align}
with subterms,
\begingroup
\allowdisplaybreaks
\begin{align*}
    \nu_5^{(0)}&= \frac{1}{1024},
    \\
    \nu_5^{(1)}&=\frac{m\chi}{8}\left(\frac{3}{16} - s^2 \right),
    \\
    \nu_5^{(2)}&= -\frac{65}{512} + \frac{3 s^2}{4}\left(\frac{3}{4} - s^2 \right) + \frac{\chi^2}{2} \left[\frac{17}{256} - \frac{s^2}{2}\left(\frac{5}{4} + {3 s^2}\right) + {m^2} \left(\frac{1}{16} - {3 s^2}\right)\right],
    \\
    \nu_5^{(3)}&= -\frac{m\chi}{4}\left\{\frac{5}{8}\left(7 - 72 s^2\right) +{\chi^2}\left[{3}{8}\left(3 - 8 s^2\right)-{5 m^2}\right]\right\} ,
    \\
    \nu_5^{(4)} &= \frac{105}{512} \left(5 - 96 s^2\right) + \frac{\chi^2}{32} \left[\frac{5}{8}\left(119 + 96 s^2\right)-{945 m^2}\right] + \frac{\chi^4}{32} \left[\frac{1}{16}\left(13 - 96 s^2\right)-{15 m^2}\right],
    \\
    \nu_5^{(5)} &= \frac{m\chi}{32}\left(\frac{9009}{2} - 189 \chi^2 + \frac{9 \chi^4}{2}\right),
    \\
    \nu_5^{(6)} &= -\frac{1}{256}\left(51051 - 9009 \chi^2 - 63 \chi^4 + 5 \chi^6\right),
\end{align*}
and,
\begin{align*}
    \nu_7^{(0)}&= -\frac{5}{32768},
    \\
    \nu_7^{(1)}&= -\frac{m\chi}{64} \left(\frac{5}{16}-\frac{5261 s^2}{3598}-\frac{85 s^4}{1799}+\frac{17 s^6}{1799}\right),
    \\
    \nu_7^{(2)}&= -\bigg\{\frac{51}{4096}-\frac{51 s^2}{256}+\frac{15 s^4}{32} 
    \\ 
    & \qquad - \frac{\chi^2}{2}\left[\frac{131}{2048}-\frac{63 s^2}{128}+\frac{15 s^4}{16} - m^2 \left( \frac{93}{256}+\frac{22675 s^2}{43176}+\frac{31705 s^4}{10794}+\frac{4925 s^6}{5397} \right) \right]\bigg\},
    \\
    \nu_7^{(3)}&= \frac{5m\chi}{2}\bigg\{\frac{125}{256}+\frac{3809 s^2}{28784}+\frac{25279 s^4}{14392}+\frac{7897 s^6}{14392}
    \\
    &
    \qquad 
    -\chi^2\left[\frac{17}{256}+\frac{5745 s^2}{28784}-\frac{8313 s^4}{14392}-\frac{2655s^6}{14392}-m^2 \left(\frac{3}{16}+\frac{5261 s^2}{3598}+\frac{85 s^4}{1799}-\frac{17s^6}{1799}\right)\right] \bigg\},
    \\
    \nu_7^{(4)} &= -\frac{5}{4}\bigg\{ \frac{5481}{2048}-\frac{105 s^2}{32}+\frac{105 s^4}{8}
    \\
    & \qquad -\chi^2 \left[\frac{1481}{1024}+\frac{51
   s^2}{16}+\frac{9 s^4}{4}-m^2 \left(\frac{245}{64}+\frac{26305 s^2}{514}+\frac{425
   s^4}{257}-\frac{85 s^6}{257}\right)\right]
    \\
    & \qquad \qquad 
    +\chi^4 \left[\frac{653}{10240}-\frac{17 s^2}{32}+\frac{9 s^4}{8}-m^2 \left(\frac{53}{64}-\frac{5261s^2}{3598}-\frac{85 s^4}{1799}+\frac{17 s^6}{1799}\right) +\frac{35m^4}{16}\right]\bigg\},
    \\
    \nu_7^{(5)} &= \frac{5m\chi}{16}\bigg[\frac{3003}{32}+\frac{868065 s^2}{1028}+\frac{14025 s^4}{514}-\frac{2805 s^6}{514}
    \\
    & \qquad \qquad  \qquad 
    -\chi^2\left(\frac{1463}{16}-\frac{693 m^2}{2}+\frac{26305 s^2}{514}+\frac{425 s^4}{257}-\frac{85s^6}{257}\right)
    \\
    & \qquad \qquad \qquad \qquad \qquad \qquad 
    -\chi^4 \left(\frac{13}{32}-\frac{7 m^2}{2}-\frac{15783 s^2}{7196}-\frac{255s^4}{3598}+\frac{51 s^6}{3598}\right)\bigg],
    \\
    \nu_7^{(6)} &= -\frac{5}{128}\bigg[\frac{27027}{16}+9009 s^2
    -\chi^2 \left(\frac{33033}{16}-\frac{45045 m^2}{2}+2079 s^2\right)
    \\
    &  \qquad \qquad  \qquad 
    +\chi^4 \left(\frac{889}{16}-693 m^2-21 s^2\right)-a^6 \left(\frac{3}{16}-\frac{21 m^2}{2}-3
   s^2\right) \bigg],
    \\
    \nu_7^{(7)} &= \frac{5m\chi}{256}\left(138567-19305 \chi^2-99 \chi^4+5 \chi^6 \right),
    \\
    \nu_7^{(8)} &= -\frac{5}{4096}\left(\frac{9561123}{4}-692835 \chi ^2+\frac{19305 \chi ^4}{2}-99 \chi ^6+\frac{35 \chi ^8}{4}\right).
\end{align*}
\endgroup
We can also extrapolate the leading-order behavior of even higher-order coefficients by first recognizing that in the $\omega \rightarrow 0$ limit, we also have $\cos2\pi\nu \rightarrow 1$ or $\nu \rightarrow l$ [along with with $\lambda_C \rightarrow l(l+1)$]. Thus, our expansion must have the following behavior in the zero-frequency limit,
\begin{align}
    \sqrt{\lambda_C} + \frac{1}{2} + \sum_{n=1}^{n_\mathrm{trunc}-1} \nu_{2n-1} {\lambda_C}^{(-2n+1)/2} &\rightarrow l + \frac{1}{2} + O\left(l^{-2n_\mathrm{trunc}+1}\right), & (\omega &\rightarrow 0).
\end{align}
By expanding the lefthand side as an asymptotic series in $l$, and requiring that all terms $O(l^{-1})$ and higher vanish, we can extract the static (zero-frequency) contribution to the higher-order terms. For example, we have,
\begin{subequations}
    \begin{align}
        \nu_1 &= \frac{1}{8} + O(\omega),
        &
        \nu_3 &= -\frac{1}{128} + O(\omega),
        \\
        \nu_5 &= \frac{1}{1024} + O(\omega),
        &
        \nu_7 &= - \frac{5}{32768} + O(\omega),
        \\
        \nu_9 &= \frac{7}{262144} + O(\omega),
        &
        \nu_{11} &= - \frac{21}{4194304} + O(\omega).
    \end{align}  
\end{subequations}

\bibliography{parent}

\begin{thebibliography}{33}%
\makeatletter
\providecommand \@ifxundefined [1]{%
 \@ifx{#1\undefined}
}%
\providecommand \@ifnum [1]{%
 \ifnum #1\expandafter \@firstoftwo
 \else \expandafter \@secondoftwo
 \fi
}%
\providecommand \@ifx [1]{%
 \ifx #1\expandafter \@firstoftwo
 \else \expandafter \@secondoftwo
 \fi
}%
\providecommand \natexlab [1]{#1}%
\providecommand \enquote  [1]{``#1''}%
\providecommand \bibnamefont  [1]{#1}%
\providecommand \bibfnamefont [1]{#1}%
\providecommand \citenamefont [1]{#1}%
\providecommand \href@noop [0]{\@secondoftwo}%
\providecommand \href [0]{\begingroup \@sanitize@url \@href}%
\providecommand \@href[1]{\@@startlink{#1}\@@href}%
\providecommand \@@href[1]{\endgroup#1\@@endlink}%
\providecommand \@sanitize@url [0]{\catcode `\\12\catcode `\$12\catcode `\&12\catcode `\#12\catcode `\^12\catcode `\_12\catcode `\%12\relax}%
\providecommand \@@startlink[1]{}%
\providecommand \@@endlink[0]{}%
\providecommand \url  [0]{\begingroup\@sanitize@url \@url }%
\providecommand \@url [1]{\endgroup\@href {#1}{\urlprefix }}%
\providecommand \urlprefix  [0]{URL }%
\providecommand \Eprint [0]{\href }%
\providecommand \doibase [0]{https://doi.org/}%
\providecommand \selectlanguage [0]{\@gobble}%
\providecommand \bibinfo  [0]{\@secondoftwo}%
\providecommand \bibfield  [0]{\@secondoftwo}%
\providecommand \translation [1]{[#1]}%
\providecommand \BibitemOpen [0]{}%
\providecommand \bibitemStop [0]{}%
\providecommand \bibitemNoStop [0]{.\EOS\space}%
\providecommand \EOS [0]{\spacefactor3000\relax}%
\providecommand \BibitemShut  [1]{\csname bibitem#1\endcsname}%
\let\auto@bib@innerbib\@empty
\bibitem [{\citenamefont {Teukolsky}(1973)}]{Teuk73}%
  \BibitemOpen
  \bibfield  {author} {\bibinfo {author} {\bibfnamefont {S.}~\bibnamefont {Teukolsky}},\ }\bibfield  {title} {\bibinfo {title} {{Perturbations of a rotating black hole. I. Fundamental equations for gravitational, electromagnetic, and neutrino-field perturbations}},\ }\href@noop {} {\bibfield  {journal} {\bibinfo  {journal} {Astrophys. J.}\ }\textbf {\bibinfo {volume} {185}},\ \bibinfo {pages} {635} (\bibinfo {year} {1973})}\BibitemShut {NoStop}%
\bibitem [{\citenamefont {{Brill}}\ \emph {et~al.}(1972)\citenamefont {{Brill}}, \citenamefont {{Chrzanowski}}, \citenamefont {{Pereira}}, \citenamefont {{Fackerell}},\ and\ \citenamefont {{Ipser}}}]{BrilETC72}%
  \BibitemOpen
  \bibfield  {author} {\bibinfo {author} {\bibfnamefont {D.~R.}\ \bibnamefont {{Brill}}}, \bibinfo {author} {\bibfnamefont {P.~L.}\ \bibnamefont {{Chrzanowski}}}, \bibinfo {author} {\bibfnamefont {C.~M.}\ \bibnamefont {{Pereira}}}, \bibinfo {author} {\bibfnamefont {E.~D.}\ \bibnamefont {{Fackerell}}},\ and\ \bibinfo {author} {\bibfnamefont {J.~R.}\ \bibnamefont {{Ipser}}},\ }\bibfield  {title} {\bibinfo {title} {{Solution of the Scalar Wave Equation in a Kerr Background by Separation of Variables}},\ }\href {https://doi.org/10.1103/PhysRevD.5.1913} {\bibfield  {journal} {\bibinfo  {journal} {Phys. Rev. D}\ }\textbf {\bibinfo {volume} {5}},\ \bibinfo {pages} {1913} (\bibinfo {year} {1972})}\BibitemShut {NoStop}%
\bibitem [{\citenamefont {Teukolsky}(1972)}]{Teuk72}%
  \BibitemOpen
  \bibfield  {author} {\bibinfo {author} {\bibfnamefont {S.~A.}\ \bibnamefont {Teukolsky}},\ }\bibfield  {title} {\bibinfo {title} {Rotating black holes: Separable wave equations for gravitational and electromagnetic perturbations},\ }\href {https://doi.org/10.1103/PhysRevLett.29.1114} {\bibfield  {journal} {\bibinfo  {journal} {Phys. Rev. Lett.}\ }\textbf {\bibinfo {volume} {29}},\ \bibinfo {pages} {1114} (\bibinfo {year} {1972})}\BibitemShut {NoStop}%
\bibitem [{\citenamefont {{Leaver}}(1986)}]{Leav86}%
  \BibitemOpen
  \bibfield  {author} {\bibinfo {author} {\bibfnamefont {E.~W.}\ \bibnamefont {{Leaver}}},\ }\bibfield  {title} {\bibinfo {title} {{Solutions to a generalized spheroidal wave equation: Teukolsky's equations in general relativity, and the two-center problem in molecular quantum mechanics}},\ }\href {https://doi.org/10.1063/1.527130} {\bibfield  {journal} {\bibinfo  {journal} {Journal of Mathematical Physics}\ }\textbf {\bibinfo {volume} {27}},\ \bibinfo {pages} {1238} (\bibinfo {year} {1986})}\BibitemShut {NoStop}%
\bibitem [{\citenamefont {{Hughes}}(2000{\natexlab{a}})}]{Hugh00b}%
  \BibitemOpen
  \bibfield  {author} {\bibinfo {author} {\bibfnamefont {S.~A.}\ \bibnamefont {{Hughes}}},\ }\bibfield  {title} {\bibinfo {title} {{Evolution of circular, nonequatorial orbits of Kerr black holes due to gravitational-wave emission}},\ }\href {https://doi.org/10.1103/PhysRevD.61.084004} {\bibfield  {journal} {\bibinfo  {journal} {Phys. Rev. D}\ }\textbf {\bibinfo {volume} {61}},\ \bibinfo {eid} {084004} (\bibinfo {year} {2000}{\natexlab{a}})},\ \Eprint {https://arxiv.org/abs/gr-qc/9910091} {gr-qc/9910091} \BibitemShut {NoStop}%
\bibitem [{\citenamefont {Wardell}\ \emph {et~al.}(2023{\natexlab{a}})\citenamefont {Wardell}, \citenamefont {Warburton}, \citenamefont {Fransen}, \citenamefont {Upton}, \citenamefont {Cunningham}, \citenamefont {Ottewill},\ and\ \citenamefont {Casals}}]{BHPT_SWSH}%
  \BibitemOpen
  \bibfield  {author} {\bibinfo {author} {\bibfnamefont {B.}~\bibnamefont {Wardell}}, \bibinfo {author} {\bibfnamefont {N.}~\bibnamefont {Warburton}}, \bibinfo {author} {\bibfnamefont {K.}~\bibnamefont {Fransen}}, \bibinfo {author} {\bibfnamefont {S.}~\bibnamefont {Upton}}, \bibinfo {author} {\bibfnamefont {K.}~\bibnamefont {Cunningham}}, \bibinfo {author} {\bibfnamefont {A.}~\bibnamefont {Ottewill}},\ and\ \bibinfo {author} {\bibfnamefont {M.}~\bibnamefont {Casals}},\ }\href {https://doi.org/10.5281/zenodo.8112931} {\bibinfo {title} {Spinweightedspheroidalharmonics}} (\bibinfo {year} {2023}{\natexlab{a}})\BibitemShut {NoStop}%
\bibitem [{\citenamefont {Gourgoulhon}\ \emph {et~al.}()\citenamefont {Gourgoulhon}, \citenamefont {Tiec}, \citenamefont {Vincent},\ and\ \citenamefont {Warburton}}]{BHPT_KGW}%
  \BibitemOpen
  \bibfield  {author} {\bibinfo {author} {\bibfnamefont {E.}~\bibnamefont {Gourgoulhon}}, \bibinfo {author} {\bibfnamefont {A.~L.}\ \bibnamefont {Tiec}}, \bibinfo {author} {\bibfnamefont {F.}~\bibnamefont {Vincent}},\ and\ \bibinfo {author} {\bibfnamefont {N.}~\bibnamefont {Warburton}},\ }\href {https://github.com/BlackHolePerturbationToolkit/kerrgeodesic_gw} {\bibinfo {title} {{BlackHolePerturbationToolkit/kerrgeodesic{\_}gw}}}\BibitemShut {NoStop}%
\bibitem [{\citenamefont {Park}(2023)}]{Park_SPH}%
  \BibitemOpen
  \bibfield  {author} {\bibinfo {author} {\bibfnamefont {S.}~\bibnamefont {Park}},\ }\href {https://doi.org/10.5281/zenodo.10426098} {\bibinfo {title} {syp2001/spheroidal: spheroidal v0.1.1}} (\bibinfo {year} {2023})\BibitemShut {NoStop}%
\bibitem [{\citenamefont {Lo}(2023)}]{Lo_SWSH}%
  \BibitemOpen
  \bibfield  {author} {\bibinfo {author} {\bibfnamefont {R.~K.~L.}\ \bibnamefont {Lo}},\ }\href {https://github.com/ricokaloklo/SpinWeightedSpheroidalHarmonics.jl} {\bibinfo {title} {ricokaloklo/spinweightedspheroidalharmonics}} (\bibinfo {year} {2023})\BibitemShut {NoStop}%
\bibitem [{\citenamefont {Sasaki}\ and\ \citenamefont {Nakamura}(1982)}]{SasaNaka82}%
  \BibitemOpen
  \bibfield  {author} {\bibinfo {author} {\bibfnamefont {M.}~\bibnamefont {Sasaki}}\ and\ \bibinfo {author} {\bibfnamefont {T.}~\bibnamefont {Nakamura}},\ }\bibfield  {title} {\bibinfo {title} {A class of new perturbation equations for the kerr geometry},\ }\href {https://doi.org/http://dx.doi.org/10.1016/0375-9601(82)90507-2} {\bibfield  {journal} {\bibinfo  {journal} {Physics Letters A}\ }\textbf {\bibinfo {volume} {89}},\ \bibinfo {pages} {68 } (\bibinfo {year} {1982})}\BibitemShut {NoStop}%
\bibitem [{\citenamefont {{Mano}}\ \emph {et~al.}(1996)\citenamefont {{Mano}}, \citenamefont {{Suzuki}},\ and\ \citenamefont {{Takasugi}}}]{ManoSuzuTaka96b}%
  \BibitemOpen
  \bibfield  {author} {\bibinfo {author} {\bibfnamefont {S.}~\bibnamefont {{Mano}}}, \bibinfo {author} {\bibfnamefont {H.}~\bibnamefont {{Suzuki}}},\ and\ \bibinfo {author} {\bibfnamefont {E.}~\bibnamefont {{Takasugi}}},\ }\bibfield  {title} {\bibinfo {title} {{Analytic Solutions of the Teukolsky Equation and Their Low Frequency Expansions}},\ }\href {https://doi.org/10.1143/PTP.95.1079} {\bibfield  {journal} {\bibinfo  {journal} {Progress of Theoretical Physics}\ }\textbf {\bibinfo {volume} {95}},\ \bibinfo {pages} {1079} (\bibinfo {year} {1996})},\ \Eprint {https://arxiv.org/abs/gr-qc/9603020} {gr-qc/9603020} \BibitemShut {NoStop}%
\bibitem [{\citenamefont {{Hughes}}(2000{\natexlab{b}})}]{Hugh00}%
  \BibitemOpen
  \bibfield  {author} {\bibinfo {author} {\bibfnamefont {S.~A.}\ \bibnamefont {{Hughes}}},\ }\bibfield  {title} {\bibinfo {title} {{Computing radiation from Kerr black holes: Generalization of the Sasaki-Nakamura equation}},\ }\href {https://doi.org/10.1103/PhysRevD.62.044029} {\bibfield  {journal} {\bibinfo  {journal} {Phys. Rev. D}\ }\textbf {\bibinfo {volume} {62}},\ \bibinfo {eid} {044029} (\bibinfo {year} {2000}{\natexlab{b}})},\ \Eprint {https://arxiv.org/abs/gr-qc/0002043} {gr-qc/0002043} \BibitemShut {NoStop}%
\bibitem [{\citenamefont {{Fujita}}\ and\ \citenamefont {{Tagoshi}}(2004)}]{FujiTago04}%
  \BibitemOpen
  \bibfield  {author} {\bibinfo {author} {\bibfnamefont {R.}~\bibnamefont {{Fujita}}}\ and\ \bibinfo {author} {\bibfnamefont {H.}~\bibnamefont {{Tagoshi}}},\ }\bibfield  {title} {\bibinfo {title} {{New Numerical Methods to Evaluate Homogeneous Solutions of the Teukolsky Equation}},\ }\href {https://doi.org/10.1143/PTP.112.415} {\bibfield  {journal} {\bibinfo  {journal} {Progress of Theoretical Physics}\ }\textbf {\bibinfo {volume} {112}},\ \bibinfo {pages} {415} (\bibinfo {year} {2004})},\ \Eprint {https://arxiv.org/abs/gr-qc/0410018} {gr-qc/0410018} \BibitemShut {NoStop}%
\bibitem [{\citenamefont {{Fujita}}\ and\ \citenamefont {{Tagoshi}}(2005)}]{FujiTago05}%
  \BibitemOpen
  \bibfield  {author} {\bibinfo {author} {\bibfnamefont {R.}~\bibnamefont {{Fujita}}}\ and\ \bibinfo {author} {\bibfnamefont {H.}~\bibnamefont {{Tagoshi}}},\ }\bibfield  {title} {\bibinfo {title} {{New Numerical Methods to Evaluate Homogeneous Solutions of the Teukolsky Equation. II ---Solutions of the Continued Fraction Equation ---}},\ }\href {https://doi.org/10.1143/PTP.113.1165} {\bibfield  {journal} {\bibinfo  {journal} {Progress of Theoretical Physics}\ }\textbf {\bibinfo {volume} {113}},\ \bibinfo {pages} {1165} (\bibinfo {year} {2005})},\ \Eprint {https://arxiv.org/abs/0904.3818} {arXiv:0904.3818 [gr-qc]} \BibitemShut {NoStop}%
\bibitem [{\citenamefont {{Sasaki}}\ and\ \citenamefont {{Tagoshi}}(2003)}]{SasaTago03}%
  \BibitemOpen
  \bibfield  {author} {\bibinfo {author} {\bibfnamefont {M.}~\bibnamefont {{Sasaki}}}\ and\ \bibinfo {author} {\bibfnamefont {H.}~\bibnamefont {{Tagoshi}}},\ }\bibfield  {title} {\bibinfo {title} {{Analytic Black Hole Perturbation Approach to Gravitational Radiation}},\ }\href {https://doi.org/10.12942/lrr-2003-6} {\bibfield  {journal} {\bibinfo  {journal} {Living Reviews in Relativity}\ }\textbf {\bibinfo {volume} {6}},\ \bibinfo {pages} {6} (\bibinfo {year} {2003})},\ \Eprint {https://arxiv.org/abs/gr-qc/0306120} {gr-qc/0306120} \BibitemShut {NoStop}%
\bibitem [{\citenamefont {Throwe}(2010)}]{Thro10}%
  \BibitemOpen
  \bibfield  {author} {\bibinfo {author} {\bibfnamefont {W.}~\bibnamefont {Throwe}},\ }\href@noop {} {\bibinfo {title} {High precision calculation of generic extreme mass ratio inspirals}} (\bibinfo {year} {2010}),\ \bibinfo {note} {http://hdl.handle.net/1721.1/61270}\BibitemShut {NoStop}%
\bibitem [{\citenamefont {Wardell}\ \emph {et~al.}(2023{\natexlab{b}})\citenamefont {Wardell}, \citenamefont {Warburton}, \citenamefont {Cunningham}, \citenamefont {Durkan}, \citenamefont {Leather}, \citenamefont {Nasipak}, \citenamefont {Kavanagh}, \citenamefont {Torres}, \citenamefont {Ottewill},\ and\ \citenamefont {Casals}}]{BHPT_TEUK}%
  \BibitemOpen
  \bibfield  {author} {\bibinfo {author} {\bibfnamefont {B.}~\bibnamefont {Wardell}}, \bibinfo {author} {\bibfnamefont {N.}~\bibnamefont {Warburton}}, \bibinfo {author} {\bibfnamefont {K.}~\bibnamefont {Cunningham}}, \bibinfo {author} {\bibfnamefont {L.}~\bibnamefont {Durkan}}, \bibinfo {author} {\bibfnamefont {B.}~\bibnamefont {Leather}}, \bibinfo {author} {\bibfnamefont {Z.}~\bibnamefont {Nasipak}}, \bibinfo {author} {\bibfnamefont {C.}~\bibnamefont {Kavanagh}}, \bibinfo {author} {\bibfnamefont {T.}~\bibnamefont {Torres}}, \bibinfo {author} {\bibfnamefont {A.}~\bibnamefont {Ottewill}},\ and\ \bibinfo {author} {\bibfnamefont {M.}~\bibnamefont {Casals}},\ }\href {https://doi.org/10.5281/zenodo.10040501} {\bibinfo {title} {Teukolsky}} (\bibinfo {year} {2023}{\natexlab{b}})\BibitemShut {NoStop}%
\bibitem [{\citenamefont {{Castro}}\ \emph {et~al.}(2013{\natexlab{a}})\citenamefont {{Castro}}, \citenamefont {{Lapan}}, \citenamefont {{Maloney}},\ and\ \citenamefont {{Rodriguez}}}]{CastETC13a}%
  \BibitemOpen
  \bibfield  {author} {\bibinfo {author} {\bibfnamefont {A.}~\bibnamefont {{Castro}}}, \bibinfo {author} {\bibfnamefont {J.~M.}\ \bibnamefont {{Lapan}}}, \bibinfo {author} {\bibfnamefont {A.}~\bibnamefont {{Maloney}}},\ and\ \bibinfo {author} {\bibfnamefont {M.~J.}\ \bibnamefont {{Rodriguez}}},\ }\bibfield  {title} {\bibinfo {title} {{Black hole monodromy and conformal field theory}},\ }\href {https://doi.org/10.1103/PhysRevD.88.044003} {\bibfield  {journal} {\bibinfo  {journal} {Phys. Rev. D}\ }\textbf {\bibinfo {volume} {88}},\ \bibinfo {eid} {044003} (\bibinfo {year} {2013}{\natexlab{a}})},\ \Eprint {https://arxiv.org/abs/1303.0759} {arXiv:1303.0759 [hep-th]} \BibitemShut {NoStop}%
\bibitem [{\citenamefont {{Castro}}\ \emph {et~al.}(2013{\natexlab{b}})\citenamefont {{Castro}}, \citenamefont {{Lapan}}, \citenamefont {{Maloney}},\ and\ \citenamefont {{Rodriguez}}}]{CastETC13b}%
  \BibitemOpen
  \bibfield  {author} {\bibinfo {author} {\bibfnamefont {A.}~\bibnamefont {{Castro}}}, \bibinfo {author} {\bibfnamefont {J.~M.}\ \bibnamefont {{Lapan}}}, \bibinfo {author} {\bibfnamefont {A.}~\bibnamefont {{Maloney}}},\ and\ \bibinfo {author} {\bibfnamefont {M.~J.}\ \bibnamefont {{Rodriguez}}},\ }\bibfield  {title} {\bibinfo {title} {{Black hole scattering from monodromy}},\ }\href {https://doi.org/10.1088/0264-9381/30/16/165005} {\bibfield  {journal} {\bibinfo  {journal} {Classical and Quantum Gravity}\ }\textbf {\bibinfo {volume} {30}},\ \bibinfo {eid} {165005} (\bibinfo {year} {2013}{\natexlab{b}})},\ \Eprint {https://arxiv.org/abs/1304.3781} {arXiv:1304.3781 [hep-th]} \BibitemShut {NoStop}%
\bibitem [{\citenamefont {{Kavanagh}}\ \emph {et~al.}(2016)\citenamefont {{Kavanagh}}, \citenamefont {{Ottewill}},\ and\ \citenamefont {{Wardell}}}]{KavaOtteWard16}%
  \BibitemOpen
  \bibfield  {author} {\bibinfo {author} {\bibfnamefont {C.}~\bibnamefont {{Kavanagh}}}, \bibinfo {author} {\bibfnamefont {A.~C.}\ \bibnamefont {{Ottewill}}},\ and\ \bibinfo {author} {\bibfnamefont {B.}~\bibnamefont {{Wardell}}},\ }\bibfield  {title} {\bibinfo {title} {{Analytical high-order post-Newtonian expansions for spinning extreme mass ratio binaries}},\ }\href {https://doi.org/10.1103/PhysRevD.93.124038} {\bibfield  {journal} {\bibinfo  {journal} {Phys. Rev. D}\ }\textbf {\bibinfo {volume} {93}},\ \bibinfo {eid} {124038} (\bibinfo {year} {2016})},\ \Eprint {https://arxiv.org/abs/1601.03394} {arXiv:1601.03394 [gr-qc]} \BibitemShut {NoStop}%
\bibitem [{\citenamefont {Casals}\ and\ \citenamefont {Zimmerman}(2019)}]{CasaZimm19}%
  \BibitemOpen
  \bibfield  {author} {\bibinfo {author} {\bibfnamefont {M.}~\bibnamefont {Casals}}\ and\ \bibinfo {author} {\bibfnamefont {P.}~\bibnamefont {Zimmerman}},\ }\bibfield  {title} {\bibinfo {title} {{Perturbations of an extremal Kerr spacetime: Analytic framework and late-time tails}},\ }\href {https://doi.org/10.1103/PhysRevD.100.124027} {\bibfield  {journal} {\bibinfo  {journal} {Phys. Rev. D}\ }\textbf {\bibinfo {volume} {100}},\ \bibinfo {pages} {124027} (\bibinfo {year} {2019})},\ \Eprint {https://arxiv.org/abs/1801.05830} {arXiv:1801.05830 [gr-qc]} \BibitemShut {NoStop}%
\bibitem [{\citenamefont {Casals}\ and\ \citenamefont {Longo~Micchi}(2019)}]{CasaMicc19}%
  \BibitemOpen
  \bibfield  {author} {\bibinfo {author} {\bibfnamefont {M.}~\bibnamefont {Casals}}\ and\ \bibinfo {author} {\bibfnamefont {L.~F.}\ \bibnamefont {Longo~Micchi}},\ }\bibfield  {title} {\bibinfo {title} {{Spectroscopy of extremal and near-extremal Kerr black holes}},\ }\href {https://doi.org/10.1103/PhysRevD.99.084047} {\bibfield  {journal} {\bibinfo  {journal} {Phys. Rev. D}\ }\textbf {\bibinfo {volume} {99}},\ \bibinfo {pages} {084047} (\bibinfo {year} {2019})},\ \Eprint {https://arxiv.org/abs/1901.04586} {arXiv:1901.04586 [gr-qc]} \BibitemShut {NoStop}%
\bibitem [{\citenamefont {Bautista}\ \emph {et~al.}(2024)\citenamefont {Bautista}, \citenamefont {Bonelli}, \citenamefont {Iossa}, \citenamefont {Tanzini},\ and\ \citenamefont {Zhou}}]{BautETC24}%
  \BibitemOpen
  \bibfield  {author} {\bibinfo {author} {\bibfnamefont {Y.~F.}\ \bibnamefont {Bautista}}, \bibinfo {author} {\bibfnamefont {G.}~\bibnamefont {Bonelli}}, \bibinfo {author} {\bibfnamefont {C.}~\bibnamefont {Iossa}}, \bibinfo {author} {\bibfnamefont {A.}~\bibnamefont {Tanzini}},\ and\ \bibinfo {author} {\bibfnamefont {Z.}~\bibnamefont {Zhou}},\ }\bibfield  {title} {\bibinfo {title} {{Black hole perturbation theory meets CFT2: Kerr-Compton amplitudes from Nekrasov-Shatashvili functions}},\ }\href {https://doi.org/10.1103/PhysRevD.109.084071} {\bibfield  {journal} {\bibinfo  {journal} {Phys. Rev. D}\ }\textbf {\bibinfo {volume} {109}},\ \bibinfo {pages} {084071} (\bibinfo {year} {2024})},\ \Eprint {https://arxiv.org/abs/2312.05965} {arXiv:2312.05965 [hep-th]} \BibitemShut {NoStop}%
\bibitem [{\citenamefont {Carneiro~da Cunha}\ and\ \citenamefont {Novaes}(2016)}]{CarnNova16}%
  \BibitemOpen
  \bibfield  {author} {\bibinfo {author} {\bibfnamefont {B.}~\bibnamefont {Carneiro~da Cunha}}\ and\ \bibinfo {author} {\bibfnamefont {F.}~\bibnamefont {Novaes}},\ }\bibfield  {title} {\bibinfo {title} {{Kerr\textendash{}de Sitter greybody factors via isomonodromy}},\ }\href {https://doi.org/10.1103/PhysRevD.93.024045} {\bibfield  {journal} {\bibinfo  {journal} {Phys. Rev. D}\ }\textbf {\bibinfo {volume} {93}},\ \bibinfo {pages} {024045} (\bibinfo {year} {2016})},\ \Eprint {https://arxiv.org/abs/1508.04046} {arXiv:1508.04046 [hep-th]} \BibitemShut {NoStop}%
\bibitem [{\citenamefont {Novaes}\ \emph {et~al.}(2019)\citenamefont {Novaes}, \citenamefont {Marinho}, \citenamefont {Lencs\'es},\ and\ \citenamefont {Casals}}]{NovaETC19}%
  \BibitemOpen
  \bibfield  {author} {\bibinfo {author} {\bibfnamefont {F.}~\bibnamefont {Novaes}}, \bibinfo {author} {\bibfnamefont {C.}~\bibnamefont {Marinho}}, \bibinfo {author} {\bibfnamefont {M.}~\bibnamefont {Lencs\'es}},\ and\ \bibinfo {author} {\bibfnamefont {M.}~\bibnamefont {Casals}},\ }\bibfield  {title} {\bibinfo {title} {{Kerr-de Sitter Quasinormal Modes via Accessory Parameter Expansion}},\ }\href {https://doi.org/10.1007/JHEP05(2019)033} {\bibfield  {journal} {\bibinfo  {journal} {JHEP}\ }\textbf {\bibinfo {volume} {2019}}\bibfield  {number} {\bibinfo  {number} { (05)},\ \bibinfo {pages} {033}},\ }\Eprint {https://arxiv.org/abs/1811.11912} {arXiv:1811.11912 [gr-qc]} \BibitemShut {NoStop}%
\bibitem [{\citenamefont {Daalhuis}\ and\ \citenamefont {Olver}(1995)}]{DaalOlve95}%
  \BibitemOpen
  \bibfield  {author} {\bibinfo {author} {\bibfnamefont {A.~B.~O.}\ \bibnamefont {Daalhuis}}\ and\ \bibinfo {author} {\bibfnamefont {F.~W.~J.}\ \bibnamefont {Olver}},\ }\bibfield  {title} {\bibinfo {title} {On the calculation of stokes multipliers for linear differential equations of the second order},\ }\href {https://doi.org/10.4310/MAA.1995.v2.n3.a6} {\bibfield  {journal} {\bibinfo  {journal} {Methods Appl.~Anal.}\ ,\ \bibinfo {pages} {348}} (\bibinfo {year} {1995})}\BibitemShut {NoStop}%
\bibitem [{\citenamefont {Misner}\ \emph {et~al.}(1973)\citenamefont {Misner}, \citenamefont {Thorne},\ and\ \citenamefont {Wheeler}}]{MisnThorWhee73}%
  \BibitemOpen
  \bibfield  {author} {\bibinfo {author} {\bibfnamefont {C.}~\bibnamefont {Misner}}, \bibinfo {author} {\bibfnamefont {K.}~\bibnamefont {Thorne}},\ and\ \bibinfo {author} {\bibfnamefont {J.}~\bibnamefont {Wheeler}},\ }\href@noop {} {\emph {\bibinfo {title} {{Gravitation}}}}\ (\bibinfo  {publisher} {Freeman},\ \bibinfo {address} {San Francisco, CA, U.S.A.},\ \bibinfo {year} {1973})\BibitemShut {NoStop}%
\bibitem [{{\relax DLMF}()}]{DLMF}%
  \BibitemOpen
  {\relax DLMF},\ \href {https://dlmf.nist.gov/} {\bibinfo {title} {{\it NIST Digital Library of Mathematical Functions}}},\ \bibinfo {howpublished} {\url{https://dlmf.nist.gov/}, Release 1.2.1 of 2024-06-15},\ \bibinfo {note} {f.~W.~J. Olver, A.~B. {Olde Daalhuis}, D.~W. Lozier, B.~I. Schneider, R.~F. Boisvert, C.~W. Clark, B.~R. Miller, B.~V. Saunders, H.~S. Cohl, and M.~A. McClain, eds.}\BibitemShut {Stop}%
\bibitem [{\citenamefont {Daalhuis}\ and\ \citenamefont {Olver}(1994)}]{DaalOlve94}%
  \BibitemOpen
  \bibfield  {author} {\bibinfo {author} {\bibfnamefont {A.~B.~O.}\ \bibnamefont {Daalhuis}}\ and\ \bibinfo {author} {\bibfnamefont {F.~W.~J.}\ \bibnamefont {Olver}},\ }\bibfield  {title} {\bibinfo {title} {Exponentially improved asymptotic solutions of ordinary differential equations. ii. irregular singularities of rank one},\ }\href {https://doi.org/10.1098/rspa.1994.0047} {\bibfield  {journal} {\bibinfo  {journal} {Proceedings of the Royal Society of London. Series A: Mathematical and Physical Sciences}\ }\textbf {\bibinfo {volume} {445}},\ \bibinfo {pages} {39} (\bibinfo {year} {1994})},\ \Eprint {https://arxiv.org/abs/https://royalsocietypublishing.org/doi/pdf/10.1098/rspa.1994.0047} {https://royalsocietypublishing.org/doi/pdf/10.1098/rspa.1994.0047} \BibitemShut {NoStop}%
\bibitem [{\citenamefont {Munna}(2020)}]{Munn20}%
  \BibitemOpen
  \bibfield  {author} {\bibinfo {author} {\bibfnamefont {C.}~\bibnamefont {Munna}},\ }\bibfield  {title} {\bibinfo {title} {{Analytic post-Newtonian expansion of the energy and angular momentum radiated to infinity by eccentric-orbit nonspinning extreme-mass-ratio inspirals to the 19th order}},\ }\href {https://doi.org/10.1103/PhysRevD.102.124001} {\bibfield  {journal} {\bibinfo  {journal} {Phys. Rev. D}\ }\textbf {\bibinfo {volume} {102}},\ \bibinfo {pages} {124001} (\bibinfo {year} {2020})},\ \Eprint {https://arxiv.org/abs/2008.10622} {arXiv:2008.10622 [gr-qc]} \BibitemShut {NoStop}%
\bibitem [{\citenamefont {Stein}(2019)}]{Stei19}%
  \BibitemOpen
  \bibfield  {author} {\bibinfo {author} {\bibfnamefont {L.~C.}\ \bibnamefont {Stein}},\ }\bibfield  {title} {\bibinfo {title} {{qnm: A Python package for calculating Kerr quasinormal modes, separation constants, and spherical-spheroidal mixing coefficients}},\ }\href {https://doi.org/10.21105/joss.01683} {\bibfield  {journal} {\bibinfo  {journal} {J. Open Source Softw.}\ }\textbf {\bibinfo {volume} {4}},\ \bibinfo {pages} {1683} (\bibinfo {year} {2019})},\ \Eprint {https://arxiv.org/abs/1908.10377} {arXiv:1908.10377 [gr-qc]} \BibitemShut {NoStop}%
\bibitem [{\citenamefont {Carneiro~da Cunha}\ and\ \citenamefont {Novaes}(2015)}]{CarnNova15}%
  \BibitemOpen
  \bibfield  {author} {\bibinfo {author} {\bibfnamefont {B.}~\bibnamefont {Carneiro~da Cunha}}\ and\ \bibinfo {author} {\bibfnamefont {F.}~\bibnamefont {Novaes}},\ }\bibfield  {title} {\bibinfo {title} {{Kerr Scattering Coefficients via Isomonodromy}},\ }\href {https://doi.org/10.1007/JHEP11(2015)144} {\bibfield  {journal} {\bibinfo  {journal} {JHEP}\ }\textbf {\bibinfo {volume} {2015}}\bibfield  {number} {\bibinfo  {number} { (11)},\ \bibinfo {pages} {144}},\ }\Eprint {https://arxiv.org/abs/1506.06588} {arXiv:1506.06588 [hep-th]} \BibitemShut {NoStop}%
\bibitem [{\citenamefont {Aminov}\ and\ \citenamefont {Arnaudo}(2024)}]{AminArna24}%
  \BibitemOpen
  \bibfield  {author} {\bibinfo {author} {\bibfnamefont {G.}~\bibnamefont {Aminov}}\ and\ \bibinfo {author} {\bibfnamefont {P.}~\bibnamefont {Arnaudo}},\ }\bibfield  {title} {\bibinfo {title} {{Black hole scattering amplitudes via analytic small-frequency expansion and monodromy}},\ }\href@noop {} {\bibfield  {journal} {\bibinfo  {journal} {ArXiv e-prints}\ } (\bibinfo {year} {2024})},\ \Eprint {https://arxiv.org/abs/2409.06681} {arXiv:2409.06681 [hep-th]} \BibitemShut {NoStop}%
\end{thebibliography}%

\end{document}